\documentclass[sn-nature, twocolumn]{sn-jnl}

\usepackage{multirow}%
\usepackage{amsmath,amssymb,amsfonts}%
\usepackage{amsthm}%
\usepackage{mathrsfs}%
\usepackage[title]{appendix}%
\usepackage{xcolor}%
\usepackage{textcomp}%
\usepackage{manyfoot}%

\usepackage{booktabs}%
\usepackage{algorithm}%
\usepackage{algorithmicx}%
\usepackage{algpseudocode}%
\usepackage{listings}
\usepackage{xcolor}
\usepackage{ifthen}

\sectionfont{\fontsize{9}{11}\selectfont}
\usepackage{ragged2e}
\justifying

\newif\ifdraft
\draftfalse                   %

\newif\ifsubmission
\submissionfalse               %

\newif\ifmonolithicfigures
\monolithicfiguresfalse  %

\ifdraft
  \newenvironment{draftBlue}{\begingroup\color{blue}}{\endgroup}
\else
  \newenvironment{draftBlue}{}{}
\fi

\ifdraft
  \PassOptionsToPackage{draft}{changes}
\else
  \PassOptionsToPackage{final}{changes}
\fi
\usepackage{changes}

\let\origincludegraphics\includegraphics
\renewcommand{\includegraphics}[2][]{%
  \ifsubmission
    \begingroup
    \setbox0=\hbox{\origincludegraphics[#1]{#2}}%
    \fbox{%
      \begin{minipage}[c][\ht0][c]{\wd0}
        \centering
        \scriptsize \textbf{[Embedded image removed for submission]}
      \end{minipage}
    }%
    \endgroup
  \else
    \origincludegraphics[#1]{#2}%
  \fi
}

\ifsubmission
  \PassOptionsToPackage{draft}{graphicx}
\else
  \PassOptionsToPackage{final}{graphicx}
\fi
\usepackage{graphicx}

\usepackage{lipsum}
\usepackage{subcaption}
\usepackage{bm}
\usepackage{enumerate}
\usepackage{makecell}
\newcommand{\res}[1]{#1}
\unboldmath
\begin{document}

\title[Article Title]{Patient-specific prediction of regional lung mechanics in ARDS patients with physics-based models: a validation study}

\author*[1]{\fnm{Maximilian} \sur{Rixner}}\email{rixner@ebenbuild.com}

\author[2]{\fnm{Maximilian} \sur{Ludwig}}

\author[3]{\fnm{Matthias} \sur{Lindner}}%

\author[3]{\fnm{Inéz} \sur{Frerichs}}

\author[3]{\fnm{Armin} \sur{Sablewski}}

\author[1]{\fnm{Karl-Robert} \sur{Wichmann}}%

\author[1]{\fnm{Max-Carl} \sur{Wachter}}%

\author[1]{\fnm{Kei W.} \sur{Müller}}%

\author[3]{\fnm{Dirk} \sur{Schädler}}%

\author[1,2]{\fnm{Wolfgang A.} \sur{Wall}}%

\author[1]{\fnm{Jonas} \sur{Biehler}}
\equalcont{These authors contributed equally to this work,shared last authorship.}

\author[3]{\fnm{Tobias} \sur{Becher}}
\equalcont{These authors contributed equally to this work, shared last authorship.}

\affil[1]{\orgname{Ebenbuild GmbH}, \orgaddress{\city{Munich}, \country{Germany}}}

\affil[2]{\orgdiv{Institute for Computational Mechanics, Department of Engineering Physics and Computation}, \orgname{TUM School of Engineering and Design, Technical University of Munich},
  \country{Germany}}

\affil[3]{\orgdiv{Department of Anesthesiology and Intensive Care Medicine}, \orgname{University Hospital Schleswig-Holstein}, \country{Germany}}

\abstract{
  Lung-protective ventilation significantly influences outcomes in ARDS patients, but identifying optimal settings remains a challenge due to pronounced inter- and intra-patient variability in lung anatomy and pathophysiology. This study demonstrates that physics-based computational lung models tailored to individual patients can predict otherwise unobservable local lung states, enabling a quantitative analysis of regional ventilation and the mechanical load experienced by lung parenchyma during ventilation. For seven mechanically ventilated ARDS patients, patient-specific computational models were generated using chest CT scan and ventilatory waveform data. By numerically resolving the lung’s interaction with ventilator-imposed pressure and flow, we predict both the regional ventilation as well as the dynamic, spatially heterogeneous states of the lung. Model-predicted ventilation distributions were validated against clinical measurements from bedside Electrical Impedance Tomography (EIT). The predicted anteroposterior ventilation profiles exhibited excellent agreement with EIT, achieving a Pearson correlation of 96\%. Across the full transverse cross-section and over the dynamic ventilation range, the models achieved an average correlation exceeding 81\% and a root mean square error below 15\%. This first systematic validation study indicates that computational lung models can reliably estimate patient-specific regional ventilation. These findings support the use of such models as a tool for individualized decision-making in mechanical ventilation, offering insights into both anatomical and functional lung characteristics that are not directly observable at the bedside. By leveraging detailed patient data and physical modeling, these models have the potential to inform more personalized and physiologically grounded ventilator settings, improving care in critically ill ARDS patients.  \\

  \textbf{New and Noteworthy:} This study presents the first systematic validation of physics-based, patient-specific computational lung models for mechanically ventilated patients. By comparing model-predicted regional ventilation with bedside Electrical Impedance Tomography, we demonstrate the ability of our computational approach to predict regional lung mechanics and patient-specific phenomena. Embedded within a computational pipeline fast enough for clinical workflows, this marks a significant step toward truly personalized, model-informed ventilation management in the ICU.
}

\vspace{1.25cm}

\maketitle

\clearpage

\section{Background}
\label{sec:introduction}

While mechanical ventilation often constitutes the only viable treatment for patients suffering from hypoxemic respiratory failure, it also poses an inherent dilemma, as subjecting the lung parenchyma to mechanical ventilation on its own may further aggrevate or even cause lung damage. This condition, known as ventilator-induced lung injury (VILI), is now understood to emerge from a complex interplay of mechanical and biological factors. Crucially, it does not stem from a single insult, but from compounding and interrelated mechanisms - most notably volutrauma, barotrauma, and atelectrauma \cite{gattinoniVentilatorInducedLung2024, serpanetoMechanicalPowerVentilation2018, silvaPowerMechanicalPower2019, batesVentilatorinducedLungInjury2018a}.
Volutrauma and barotrauma refer to damage caused by excessive stress and strain, typically resulting from overdistension of alveolar units under high volumes or pressures.
In contrast, atelectrauma arises from the repeated collapse and re-opening of unstable lung units, often at air-liquid interfaces or in regions exhibiting a heterogeneous degree of recruitment.
This cyclic process contributes to localized mechanical forces and increased energy dissipation, particularly at interfaces where open and collapsed units coexist and alveolar stability is compromised \cite{meadStressDistributionLungs1970}.
Recent experimental studies have emphasized that these localized phenomena may cause irreversible tissue damage and barrier disruption, implicating not only structural failure but also activation of mechanotransductive and inflammatory pathways \cite{gonzalez-lopezLungStrainBiological2012, gaverMechanicalVentilationEnergy2025, gabela-zunigaMicroscaleHumanizedVentilatoronachip2024}.
Other discussions of VILI have adopted rheological \cite{modestoialapontClinicalImplicationsRheological2019} or thermodynamic viewpoints, e.g., considering lung damage to arise when physiological mechanisms are no longer able to counteract the \emph{local} entropy production \cite{oliveiraEntropyProductionPressureVolume2016}.\\
What unifies each potential contributing mechanism of VILI is their spatial and temporal heterogeneity, with injurious processes being entirely inaccessible to direct clinical imaging, arising as the opaque interplay of the mechanical energy imparted by the ventilator with the heterogenous lung tissue.
These insights suggest that the severity and location of lung injury cannot be inferred from global parameters alone, such as airway pressure or mechanical power. Instead, VILI emerges from the regional distribution and temporal cycling of stress and strain, modulated by the lungs intrinsic heterogeneity. \\
It is now widely accepted that lung-protective mechanical ventilation must minimize cumulative biotrauma while simultaneously ensuring adequate gas exchange and hemodynamic stability, and recognition of VILI as a major factor contributing to high mortality and morbidity has significantly advanced acute respiratory distress syndrome (ARDS) treatment \cite{dreyfussWhatConceptVILI2016}.
In juxtaposition to the general agreement regarding the necessity of lung protective ventilation lies however the lack of any clear consensus on the \emph{specific} optimal ventilation strategy for an individual patient.
This holds particularly true in the case of ARDS, which, as a complex and clinically heterogeneous syndrome, is inherently challenging to treat, involving different phenotypes and co-morbidities.
Current standard-of-care still largely relies on broad population-average heuristics \cite{mariniEvolvingConceptsSafer2019} with relatively simple criteria, such as an acceptable range of tidal volume as a function of predicted body weight (4-8 mL/kg) \cite{grasselliESICMGuidelinesAcute2023,putensenMetaanalysisVentilationStrategies2009}.
This approach, however, fails to account for inter- and intra-patient variability \cite{kollisch-singuleLookingMacroventilatoryParameters2018} and the unresolved heterogeneous pathologies and mechanical properties of the lung tissue leading to patient-specific responses and heterogeneity of treatment effects \cite{khanPrecisionMedicineHeterogeneity2021,chenLungMorphologyImpacts2023}.
This is in line with recent clinical studies on optimal mechanical ventilation strategies for ARDS increasingly encountering inconsistent and inconclusive results as a consequence of the unresolved heterogeneity and complex pathophysiology \cite{juschtenBetweentrialHeterogeneityARDS2021} and
- unfortunately - also in line with the observation that patient outcome for ARDS has stagnated for more than two decades. Mortality remains high at roughly 40\% \cite{villarCurrentIncidenceOutcome2016, sadanaMortalityAssociatedAcute2022}.

The inability to further advance patient outcome employing general treatment protocols has led to attempts to inform ventilation strategies based on additional patient-specific information (e.g.,  EIT \cite{gomez-labergeUnifiedApproachEIT2012}),  attempts to stratify the inherent variability of ARDS into further latent subgroups and phenotypes \cite{calfeeSubphenotypesAcuteRespiratory2014, calfeeAcuteRespiratoryDistress2018, heijnenBiologicalSubphenotypesAcute2021, maddaliValidationUtilityARDS2022}, as well as general calls for the development of new, precision-guided medical approaches for lung protective ventilation \cite{gattinoniFutureMechanicalVentilation2017, beitlerAdvancingPrecisionMedicine2022a}.

In more recent developments, researchers have also proposed the use of data-driven reinforcement learning  \cite{peineDevelopmentValidationReinforcement2021, kondrupSafeMechanicalVentilation2023} for the identification of optimal ventilatory settings.
In essence, this approach mimics a highly experienced physician by looking at past treatment trajectories in order to identify optimal choices, leveraging the excellent ability of machine learning algorithms to discern patterns in large corpi of historical clinical data.
Beside lacking interpretability and explainability, it is also limited both to historical data as well as global descriptions of the patient and the lung.
In this approach, the lung itself continues to remain a black box to the treating clinician, and regional lung mechanics and heterogeneous ventilation - key factors underlying VILI - are not taken into account.

In our opinion, truly patient-specific lung-protective ventilation needs to account for the complex interplay of pressure, airflow and regional pathophysiology within the parenchyma, thereby also intrinsically necessitating a model-based approach.
Predictive models not only enable a detailed analysis of the current state of the lungs but also the evaluation of hypothetical scenarios to determine the optimal treatment for a patient without having to test different settings directly on the patient.

\begin{figure*}[hbtp]
  \centering
  \ifmonolithicfigures
    \includegraphics{JAPPL_concept_e7d.pdf}
  \else
    \includegraphics[scale=0.9]{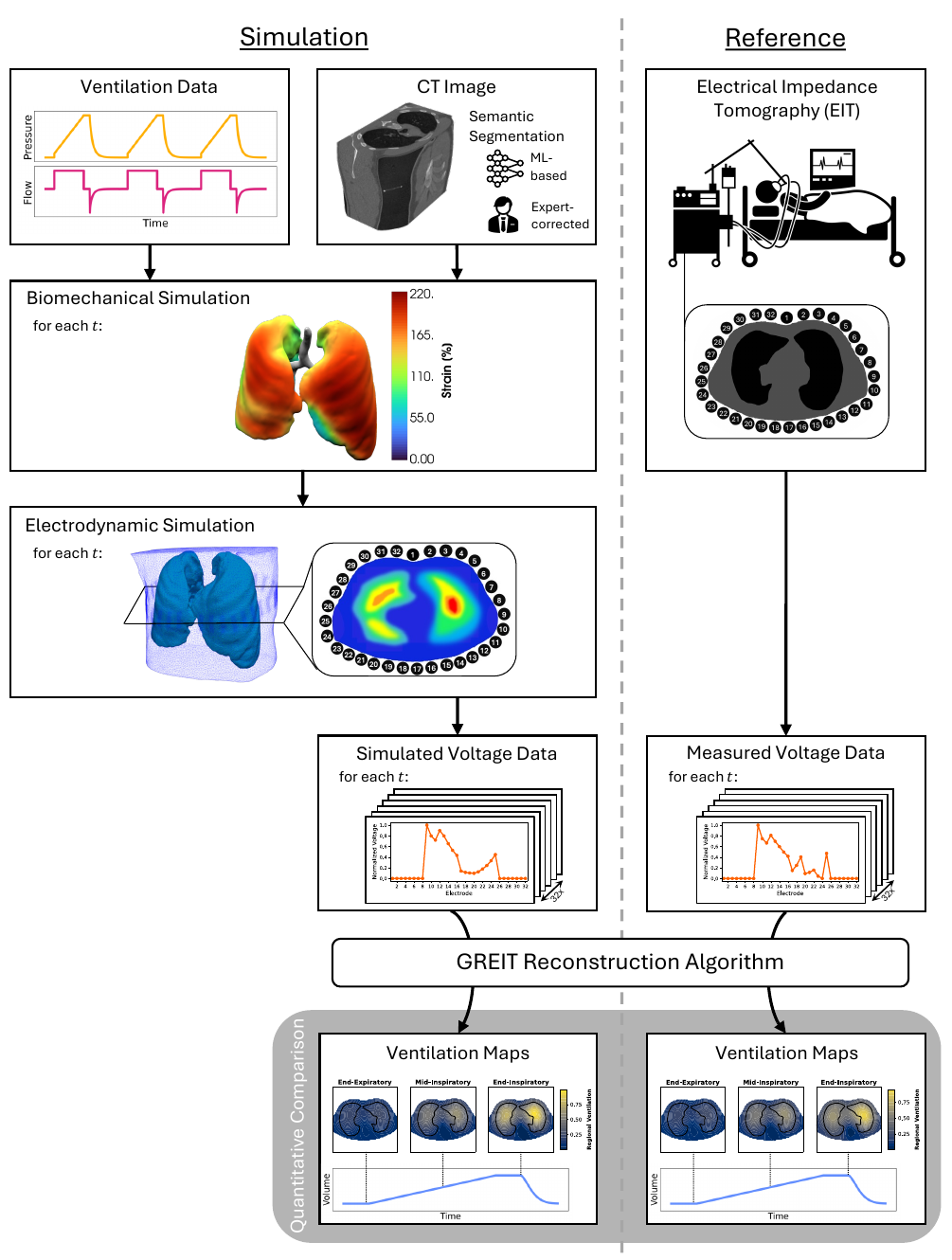}
  \fi
  \vspace{0.60cm}
  \caption{Conceptual overview - validation of computational lung model against clinical reference data. \emph{Reference:} Using an electrode belt positioned around the thorax of the patient, we can observe the regional ventilation during mechanical ventilation (EIT).  \emph{Simulation:} On the basis of a computed tomography (CT) scan we construct a patient-specific model of the lung and conduct a biomechanical and electrodynamic simulation, mimicking both the mechanical ventilation process and the propagation of electrical currents across the patient's torso. The simulated voltages obtained from the computational lung model yield a \emph{virtual} EIT,  which we compare against the bedside EIT obtained from clinical, measured voltage data for a series of timesteps $t$.}
  \label{fig:concept_illustration_validation}
\end{figure*}

While several computational lung models have been developed in the past \cite{rothComputationalModellingRespiratory2017,rothComputationalModelingRespiratory2019,neelakantanComputationalLungModelling2022}, in the context of a clinical application many models either lack the spatial resolution to resolve the heterogeneity within the lung \cite{mellenthinUsingInjuryCost2019a,reesDeterminingAppropriateModel2017,sundaresanMinimalModelLung2009,mortonVirtualPatientModel2018} or otherwise preclude personalized application due to prohibitive runtime requirements and/or complexity.
In this study we therefore pursue the validation of our own development of a reduced dimensional lung model
\cite{ismailCoupledReducedDimensional2013,rothCouplingEITComputational2017,geitnerPressureTimedependentAlveolar2024,grillSilicoHighresolutionWhole2023}, which has the unique ability to reflect patient-specific anatomy and pathophysiology, while still being able to provide results within clinically relevant time-frames, solely relying on routinely available clinical data.
In particular, we only depend on a \emph{single} static CT at a fixed pressure, as well as waveform data from the mechanical ventilator.
Making use of previously developed electrodynamic coupling \cite{rothComputationalModellingRespiratory2017, rothCouplingEITComputational2017}, the main novelty and contribution of this publication lies in the systematic evaluation and validation of model predictions against bedside Electrical Impedance Tomography (EIT), demonstrating the capability of the computational lung model to provide patient-specific, accurate predictions of regional ventilation (see Figure \ref{fig:concept_illustration_validation}).

\section{Material and Methods}
\label{sec:materials_and_methods}
This validation study assesses the accuracy of patient-specific lung models for a cohort of patients suffering from ARDS and undergoing invasive mechanical ventilation.
In order to construct and calibrate models which are able to resolve the underlying physics with a high degree of fidelity, we make use of a single, routinely available clinical chest computed tomography (CT) scan, as well as waveform data exported from the mechanical ventilator.
Afterwards, the computational lung model is used to simulate the process of mechanical ventilation, yielding predictions of regional mechanical quantities such as ventilation distribution and strain subject to intra- and interpatient variability.
Finally, we validate these predictions by comparing them against the regional ventilation distribution obtained from bedside Electrical Impedance Tomography (EIT).
While we will introduce all relevant steps in the following< sections, due to the inherent scope we must also refer the reader to aforementioned prior work covering the computational model \cite{ismailCoupledReducedDimensional2013,rothCouplingEITComputational2017,
    geitnerPressureTimedependentAlveolar2024,grillSilicoHighresolutionWhole2023} and the electrodynamic coupling \cite{rothComputationalModellingRespiratory2017, rothCouplingEITComputational2017} used for validation in full detail.
An abridged and condensed exposition of relevant details has also been included in the appendix.

\subsection{Conceptual Overview}
\label{sec:conceptual_overview}

A conceptual overview of all relevant steps underlying the model validation is depicted in Figure \ref{fig:concept_illustration_validation}.
The right column shows the established clinical process of using bedside measured electrode voltage data for pulmonary monitoring, yielding dynamic low-resolution EIT images corresponding to regional changes in electrical bioimpedance of lung tissue and, thereby, visualizing the amount of regionally contained air over time against a predefined baseline (usually end-expiratory state).
This regional ventilation distribution serves as the reference, against which we compare the model-predicted regional ventilation - see left column of Figure \ref{fig:concept_illustration_validation}.
As a first step, the patient-specific biomechanical model of the respiratory system - further described in the subsequent sections - is constructed for each study participant using a chest CT and ventilatory data, enabling the personalized prediction of the time-dependent global and local state of the lung parenchyma during mechanical ventilation.
Since the model can predict - among other quantities - regional strain and ventilation state, the biomechanical simulation also enables us to inform a subsequent electrodynamic simulation, computing the propagation of electromagnetic fields across the patient's torso.
This is possible , because the model-predicted regional ventilation state informs how the bioimpedance of the lung parenchyma is altered.
Finally, we apply the same electrical stimuli in the electrodynamic simulation as the EIT device in the patient \emph{in-vivo}, thereby obtaining \emph{simulated} voltage data, enabling in turn the reconstruction of \emph{simulated} EIT images.

In consequence, we not only predict how the forced positive-pressure ventilation interacts with the deformation of the lung parenchyma, but also how this will affect the propagation of electrodynamic currents as a function of the regional ventilation.
By essentially mirroring the entire process of ventilation and electrodynamic stimulation in the computational model, we ensure \emph{consistent} comparison of the regional ventilation arising from both model predictions and measured reference data.
The comparison of simulated and bedside EIT then defines the basis of quantitative analysis and model validation as presented in our results section.
Since the validation is intrinsically tied to EIT as available imaging modality, \emph{regional ventilation} throughout the present work will always refer to the change of local bioimpedance relative to the end-expiratory state, reported in arbitrary units (a.u.).
We emphasize that the measured voltage data from EIT is not used for model generation or calibration in any way, but exclusively serves as the reference against which we compare the model-predicted regional ventilation.

\begin{figure*}[t]
    \centering
    \ifmonolithicfigures
        \includegraphics{JAPPL_ct_slices_monolithic.pdf}
    \else
        \begin{subfigure}[b]{0.32\textwidth}
            \includegraphics[width=\textwidth]{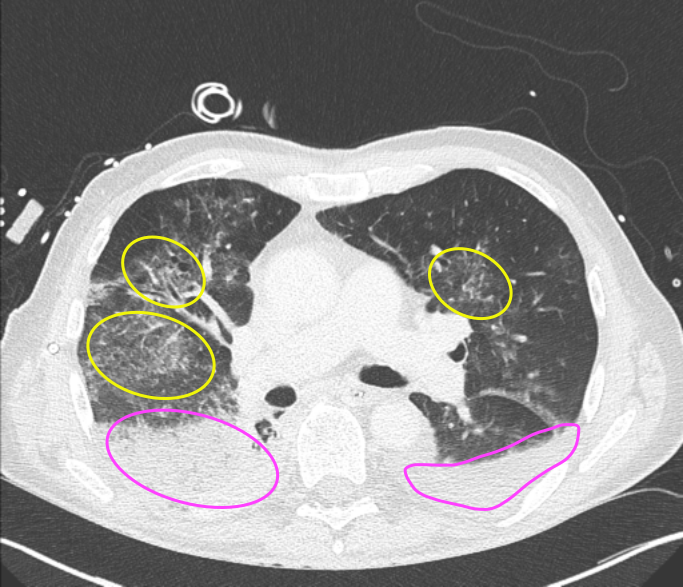}
            \caption{Patient 2}
            \label{fig:ct_slice_e2}
        \end{subfigure}
        \begin{subfigure}[b]{0.32\textwidth}
            \includegraphics[width=\textwidth]{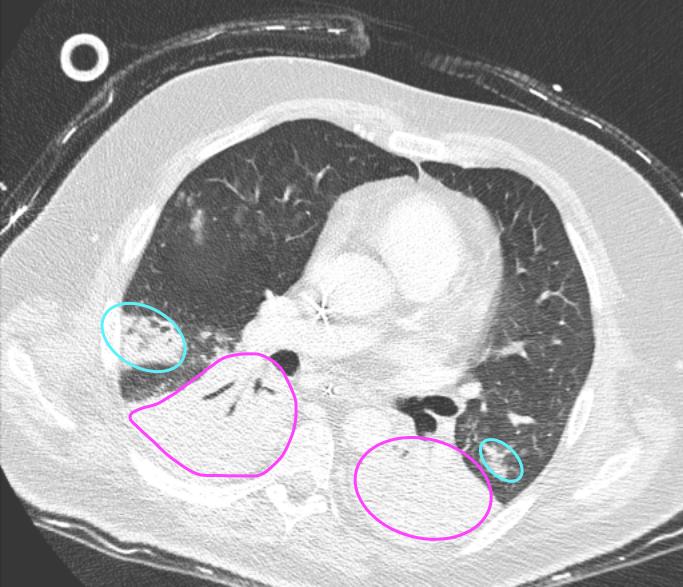}
            \caption{Patient 9}
            \label{fig:ct_slice_e9}
        \end{subfigure}
        \begin{subfigure}[b]{0.32\textwidth}
            \includegraphics[width=\textwidth]{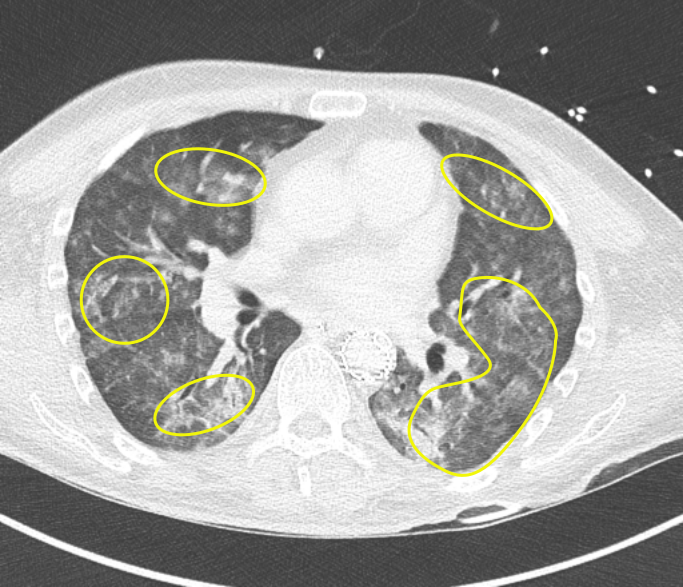}
            \caption{Patient 10}
            \label{fig:ct_slice_e10}
        \end{subfigure}
    \fi
    \caption{Exemplary transverse CT slices at the level of the EIT belt between $4th$ or $5th$ intercostal space defining the EIT reconstruction plane acquired in three of the studied patients. Markups indicate \emph{(yellow)} ground-glass opacities, \emph{(magenta)} dorsobasal atelectasis (with accompanying pleural effusion) and \emph{(cyan)} pulmonary infiltrates.}
    \label{fig:ct_slices}
\end{figure*}

\subsection{Clinical Data}
\label{sec:clinical_data}

The clinical data analyzed and used for model validation was collected within the scope of the SMART-Study (DRKS-ID: DRKS00017151). The primary purpose of that study was to validate computational models that predict not just strain but also recruitment/de-recruitment processes during recruitment maneuvers or decremental positive end-expiratory pressure (PEEP) trials, which is not within the scope of the present work.
A publication describing the entirety of the clinical dataset, as well as study and project results, is forthcoming.
Here, we only describe a subset of the data as relevant to our work in a dual-use setting, corresponding to regular pressure-controlled ventilation at fixed ventilation settings.
Briefly, ARDS patients were eligible for study inclusion if \emph{(i)} they met the Berlin definition \cite{ardsdefinitiontaskforceAcuteRespiratoryDistress2012}, \emph{(ii)} required mechanical ventilation and \emph{(iii)} a thoracic CT scan was carried out within 24 hours before their study enrolment.
Exclusion criteria were age less than 18 years, pregnancy, presence of a cardiac pacemaker or other implantable electronic devices, higher-grade burns, or skin lesions in the area of the EIT belt placement, high-grade chronic obstructive lung disease, and hemodynamic instability despite adequate fluid and catecholamine therapy.
The study was approved by the Ethics Committee of the Medical Faculty of the Christian-Albrechts-Universität Kiel and carried out in accordance with the Helsinki Declaration.
Written informed consent was obtained from all patients or their legal representatives.

\begin{table*}[ht]
    \centering
    \begin{tabular}{lcccccccccc}
        \toprule
        Dataset                    & P1   & P2   & P3   & P4   & P5   & P6   & P7   & P8   & P9   & P10  \\
        \midrule
        Age [years]                & 51   & 81   & 75   & 59   & 67   & 72   & 63   & 25   & 56   & 57   \\
        Body Height [cm]           & 155  & 180  & 175  & 168  & 184  & 180  & 172  & 180  & 186  & 190  \\
        Body Weight [kg]           & 50   & 80   & 75   & 100  & 90   & 100  & 100  & 80   & 110  & 90   \\
        Body Mass Index [kg/$m^2$] & 20.8 & 24.7 & 24.5 & 36.4 & 26.9 & 30.9 & 33.8 & 24.7 & 31.8 & 24.9 \\
        PF-Ratio [mmHg]            & 220  & 176  & 240  & 120  & 283  & 83   & 120  & 220  & 93   & 180  \\
        Tube Size [ID]             & 7    & 7.5  & 8.5  & 8    & 8    & 8.5  & 8    & 7    & 8    & 9    \\
        ASA                        & 3    & 3    & 3    & 3    & 2    & 3    & 3    & 1    & 2    & 2    \\
        Gender                     & f    & m    & m    & m    & m    & m    & f    & f    & m    & m    \\
        Phenotype of ARDS          & Pu   & Pu   & Pu   & Pu   & Ex   & Pu   & Pu   & Pu   & Pu   & Pu   \\
        \bottomrule
    \end{tabular}
    \caption{Patient cohort - phenotype abbreviations correspond to \underline{Pu}lmonary and \underline{Ex}pulmonary.}
    \label{tab:clinical_data}
\end{table*}

For each patient, a \emph{single} CT image was acquired with a Siemens Somatom or Philips IQon scanner at a pre-defined pressure state corresponding to end-expiratory lung volume (EELV) with an in plane resolution of 512x512 and a slice thickness of 2 mm according to routine clinical practice at the University Medical Center in Kiel.
For all patients, EIT examinations were carried out using the Elisa 800 (Löwenstein Medical, Bad Ems, Germany) with the textile 32-electrode belt placed around the patient’s chest circumference in one slightly oblique transverse plane at the level of the $4th$ or $5th$ intercostal space, measured in the parasternal line.
A brief overview over the patient cohort is given in Table \ref{tab:clinical_data}, and Figure \ref{fig:ct_slices} shows exemplary slices of the chest CT at the height where the EIT belt was placed. Moreover, the shown slices highlight the regional heterogeneity of the lungs with bilateral infiltrates and pulmonary edema.
In addition to the CT and EIT data, we used airway pressure, air flow, and esophageal pressure recorded during 10 breathing cycles of pressure-controlled ventilation.

\subsection{Computational Lung Model}
\label{sec:computational_lung_model}

\begin{figure*}[t]
    \centering
    \ifmonolithicfigures
        \includegraphics{JAPPL_pipeline_model_construction.pdf}
    \else
        \includegraphics[scale=0.580]{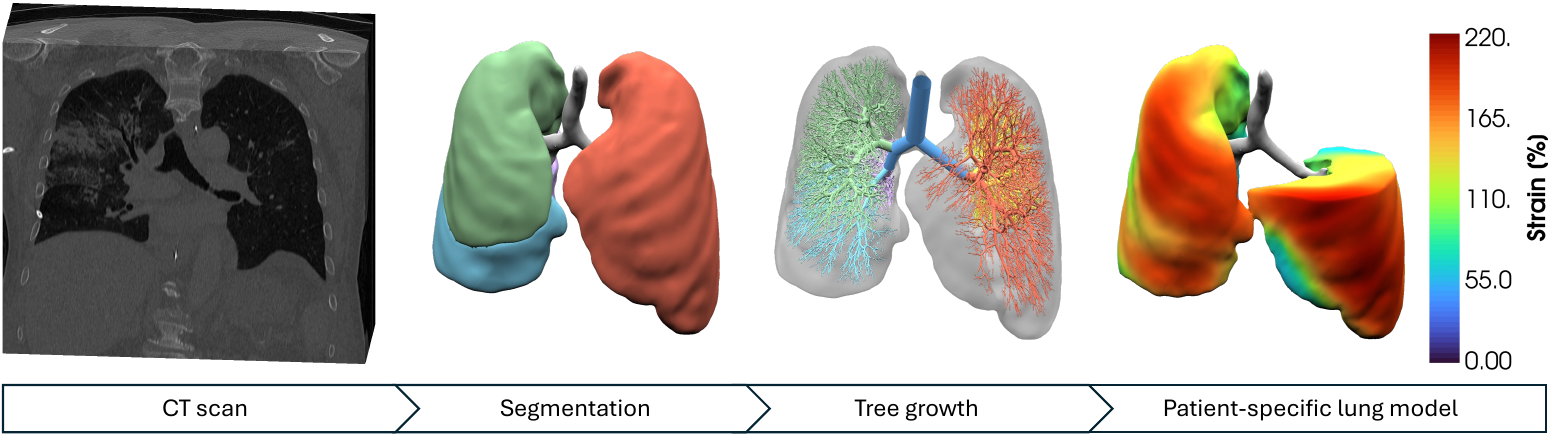}
    \fi
    \vspace{0.20cm}
    \caption{With a high-resolution CT scan as input, the generation of patient-specific models is a streamlined process starting with the segmentation of the voxel data into airways, lungs, and lobes. Beyond the resolution limitations of the CT, a morphology-based tree growth
        algorithm creates the remaining generations of conducting
        airways in a space-filling manner (for a total of $16$ generations). Finally, a patient-specific computational lung model is constructed to predict regional lung mechanics, such as the strain or regional ventilation of the lung parenchyma during mechanical ventilation.
    }
    \label{fig:pipeline_model_construction}
\end{figure*}

For each study participant, we created a patient-specific computational model capable of predicting airflow and tissue distention as well as their interaction throughout the respiratory system.
To achieve this, we first extract pertinent structures of the respiratory system from a single chest CT using computer vision and deep learning, obtaining a patient-specific anatomic representation of the lungs, lobes, and airways.\footnote{Since this study focuses on the validation of the physics-based computational model, additional manual inspection and correction of the segmentation masks was carried out to ensure absolute correctness.}
Following the approach outlined in Figure \ref{fig:pipeline_model_construction}, the patient-specific physiological features are then subsequently translated to reduced-dimensional computational elements representing the behavior of airways and the lung parenchyma, with volumetric and constitutive effects of the lung parenchyma being locally informed on the basis of Hounsfield attenuation values \cite{ismailCoupledReducedDimensional2013, rothComprehensiveComputationalHuman2017, rothCouplingEITComputational2017}.
Insights into typical constitutive behavior of lung tissue have been gained via extensive experimental studies on living lung tissue samples \cite{birzleExperimentalCharacterizationModel2018, birzleConstituentspecificMaterialBehavior2019, birzleViscoelasticNonlinearCompressible2019, birzleCoupledApproachIdentification2019}.
Further details regarding the assumed viscoelastic behavior are provided in Appendix \ref{appendix:A}.
In addition to the CT, respiratory time series data from the mechanical ventilator is used to calibrate and inform global respiratory system characteristics of the model.
Since the computational model is derived from \emph{first principle}, i.e., it is based on the description of fundamental physical laws, it allows extrapolative prediction of \emph{regional} lung mechanics and its interaction with forcing airflow and pressure (see exemplarily the strain field depicted in Figure \ref{fig:pipeline_model_construction}).
We refer the interested reader to Appendix \ref{appendix:A} for a more detailed description of the reduced dimensional computational lung model and its calibration, as well as to aforementioned prior work \cite{ismailCoupledReducedDimensional2013, rothComprehensiveComputationalHuman2017, rothCouplingEITComputational2017} for full mathematical details.
To ensure reproducibility and scalability, the entire model generation process has been automated by implementing it as in-house software packages and integrating it within an end-to-end computing pipeline. Simulations are able to execute on essentially commodity hardware with quick turnaround times. At the maximum resolution of the computational model employed for the validation results, the approximately 30 seconds of mechanical ventilation simulated correspond to less than 30 minutes of CPU time on a single core.

\begin{figure*}[t]
    \centering
    \ifmonolithicfigures
        \includegraphics{JAPPL_strain_visualization_monolithic.pdf}
    \else
        \begin{subfigure}[b]{0.26\textwidth}
            \includegraphics[width=0.90\textwidth]{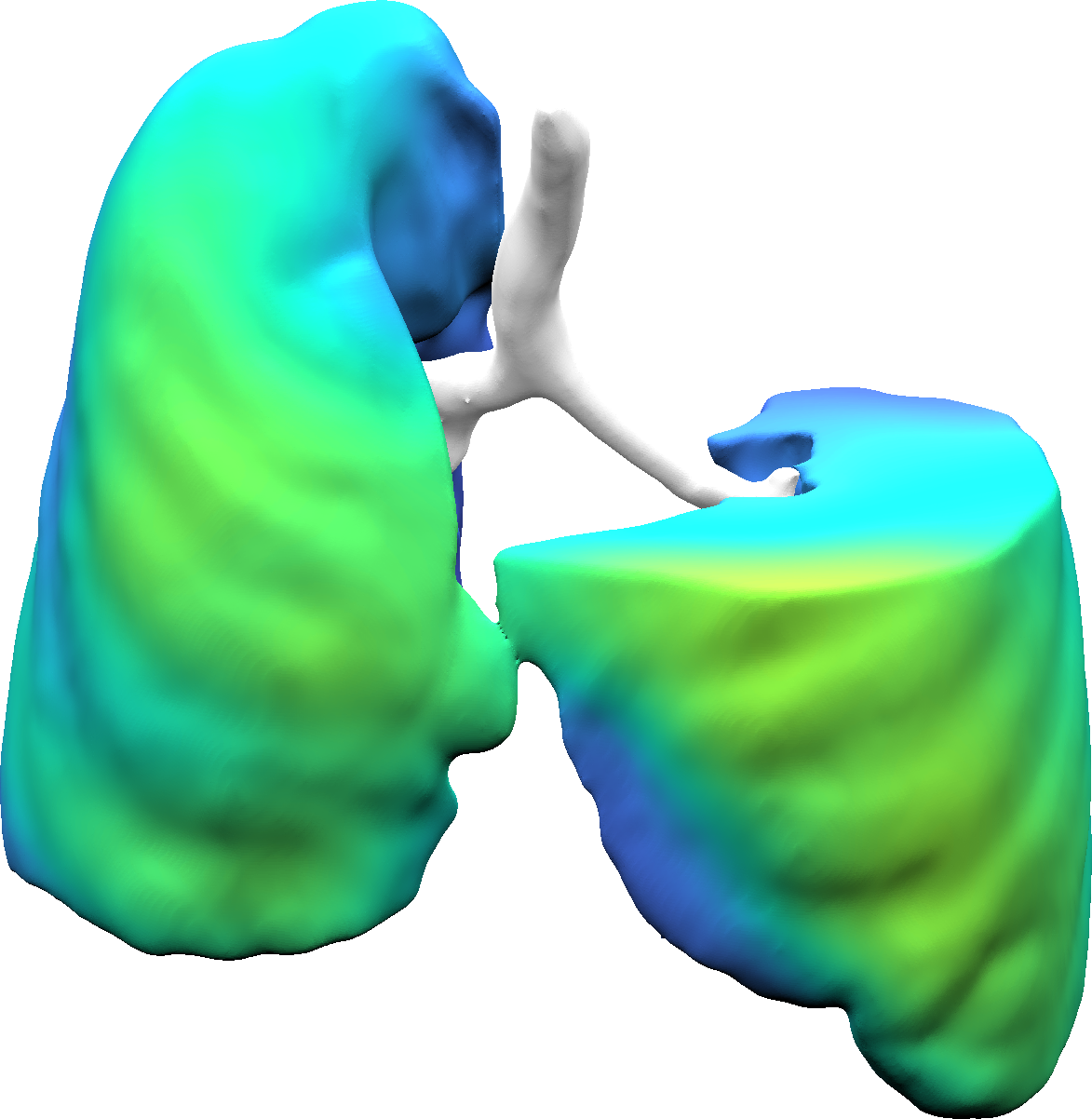}
            \vspace{0.4cm}
            \caption{Strain at p=10 cmH\textsubscript{2}O}
        \end{subfigure}
        \begin{subfigure}[b]{0.26\textwidth}
            \includegraphics[width=0.90\textwidth]{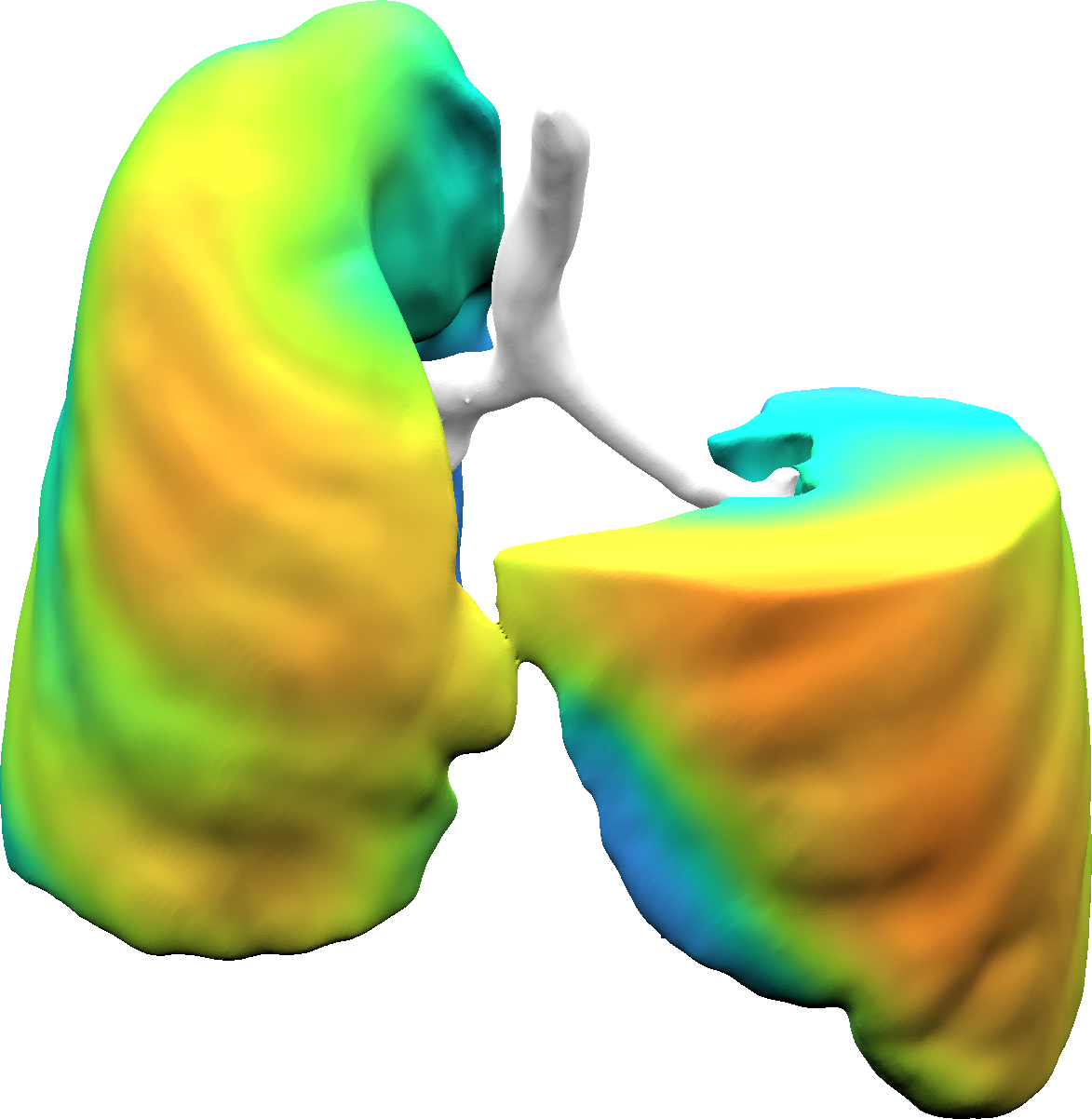}
            \vspace{0.4cm}
            \caption{Strain at p=20 cmH\textsubscript{2}O}
        \end{subfigure}
        \begin{subfigure}[b]{0.26\textwidth}
            \includegraphics[width=0.90\textwidth]{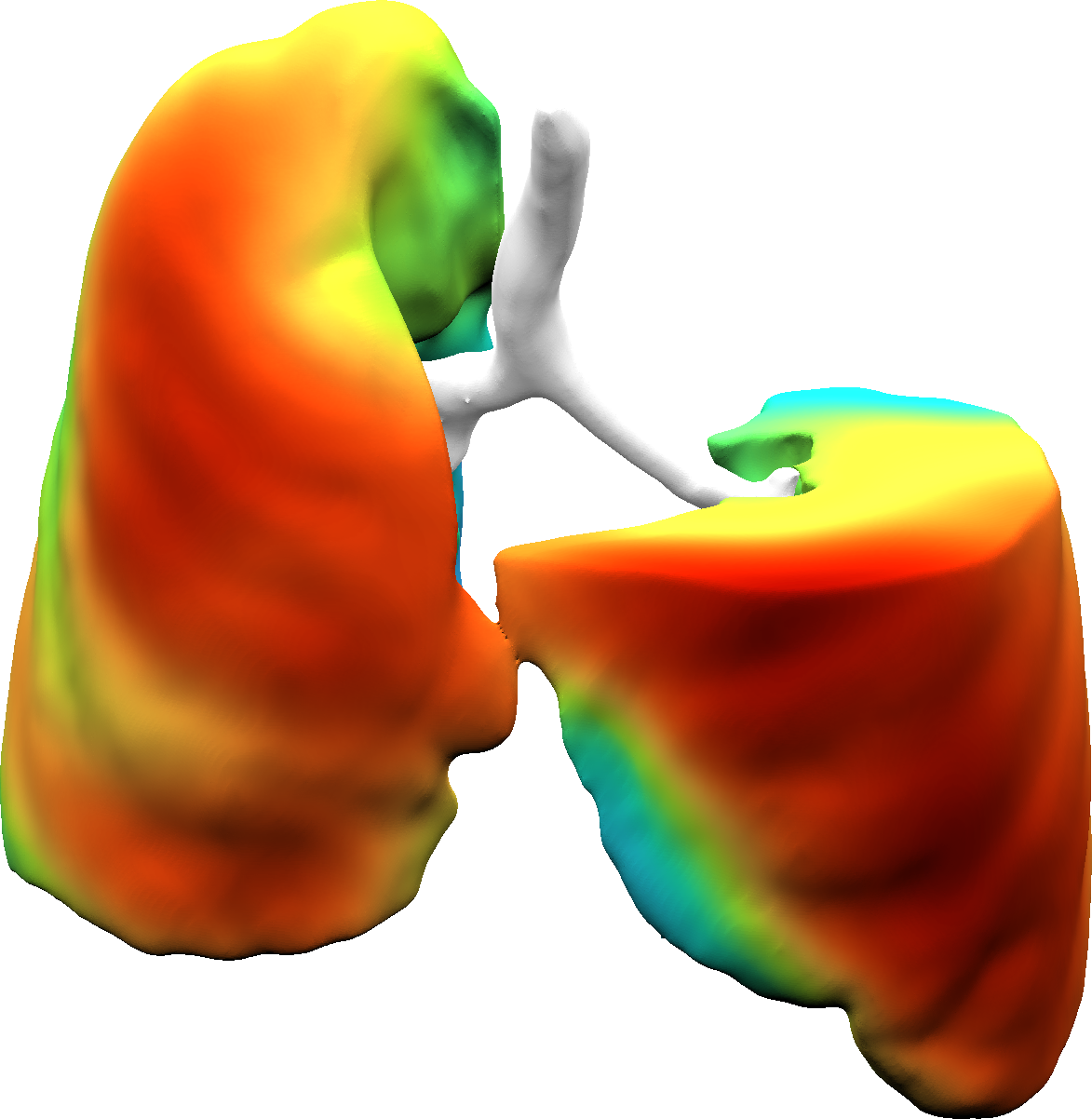}
            \vspace{0.4cm}
            \caption{Strain at p=30 cmH\textsubscript{2}O}
        \end{subfigure}
        \begin{subfigure}{0.1\textwidth}
            \vspace{-4.4cm}
            \includegraphics[width=0.90\textwidth]{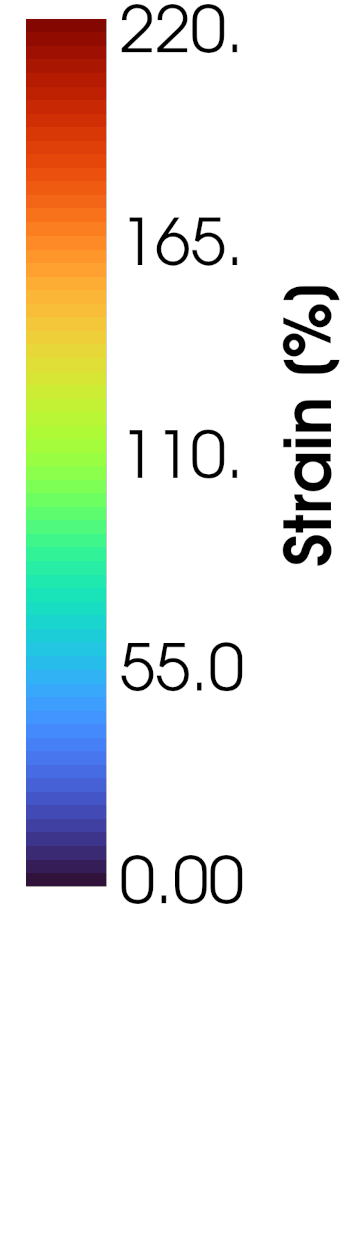}
        \end{subfigure}
    \fi
    \caption{Strain predictions for patient 2 during pressure-controlled ventilation at different points in time, showcasing the heterogeneous strain distribution within the lungs.}\label{fig:model_strain_predictions}
\end{figure*}

The resulting patient-specific, physics-based simulation models account for individual anatomy as well as pathophysiology.
In particular, the model captures the elastic interaction of the lungs with the rib cage and diaphragm and also accounts for the hydrostatic pressure resulting from gravitational load and patient positioning.
As a result, we can predict - among others - regionally resolved transient airflow, ventilation, strain, and stress over the entire lung for arbitrary - real or hypothetical - ventilation settings.
The only required input for the computational model - once constructed - is the sequence of pressure or flow values imposed by the ventilator at the endotracheal tube (depending on pressure- or volume-controlled ventilation mode), as well as patient positioning.
Exemplary model predictions, showing how the regional strain of the lung parenchyma changes with different pressure levels applied by the mechanical ventilator, are shown in Figure \ref{fig:model_strain_predictions}.

\subsection{Electrodynamic Coupling}
\label{sec:electrodynamic_simulation}

The bedside EIT carried out for each study participant yields measurements of voltage data over time during mechanical ventilation.
Using a suitable reconstruction algorithm, \emph{relative} changes of the voltage measurements with respect to an end-expiratory reference state can be used to infer corresponding \emph{relative} changes of spatially resolved  impedance within the lung parenchyma on a two-dimensional transverse plane (as defined by the electrode positions).
The deviation of impedance inferred from beside EIT in turn defines a proxy to the regional ventilation, and serves as the ground truth against which we quantitatively compare the predictions of the computational lung model.
In order to derive an equivalent representation of regional ventilation from the computational lung model, we map the model-predicted ventilation states to lung impedance changes on the basis of previously executed highly resolved three-dimensional elastodynamic reference simulations \cite{rothCorrelationAlveolarVentilation2015}.
Conditionally on known model-predicted perturbations of the parenchymal impedance, additional electrodynamic simulations can then be executed to obtain \emph{simulated} voltage data in complete equivalence to the measured clinical voltage data (see also Figure \ref{fig:concept_illustration_validation}).

Subsequently applying the identical reconstruction algorithm to the simulated voltage data yields a model-derived \emph{virtual} EIT, which is directly comparable to the clinical beside EIT.
While the model is capable of quantitative predictions across the entire lung, the reference data limits the comparison to a two-dimensional transverse plane, as well as to regional ventilation being defined in relative terms in arbitrary units (AU).
Reconstruction of regional ventilation from the measured or simulated voltage measurements is performed using GREIT (\emph{Graz consensus Reconstruction algorithm for EIT}) \cite{adlerGREITUnifiedApproach2009} both because it is as close as possible to an established standard, and also because an implementation is readily available within the open source framework EIDORS \cite{adlerEIDORSCommunitybasedExtensible2005}.
Please note that the virtual EIT and the associated regional ventilation is not inherently tied to any specific definition of strain.
Our corresponding source code for the electrodynamic simulation and the EIT reconstruction on the basis of EIDORS is available online (see availability of data and materials), and further technical details concerning EIT and the reconstruction algorithm are provided in Appendix \ref{appendix:B}. The approach follows prior work \cite{rothCouplingEITComputational2017}, providing a more in-depth and detailed description of the methodological execution.

\section{Results}
\label{sec:results}
In the following we will detail the quantitative comparison conducted between simulated and reference EIT.
Out of 10 patients enrolled in the study, one patient had to be excluded from the analysis due to loss of recorded voltage data.
Two additional patients had to be excluded because undue noise within the recorded voltage data prevented obtaining any physiological and stable reconstructions of the regional ventilation.
Thus, we conducted the assessment of the predictive capabilities of our model for 7 patients in absence of any spontaneous breathing efforts.

As a reminder, in the following we will assess and quantify the agreement of the regional ventilation as predicted by the model (simulation), comparing it against the bedside EIT (reference).
Using (\emph{s}) and (\emph{r}) to delineate between (s)imulation and (r)eference, we denote by $v^s (t)$ and $v^r(t)$ simulated and measured voltages, and refer to corresponding EIT reconstructions as $Z^{s} (t)$ and $Z^{r}(t)$.
The reconstructions $Z$ define square matrices - i.e. a grid of pixel intensity values - which represent the regional change of the impedance of the lung parenchyma relative to the end-expiratory state on a two-dimensional transverse plane.
In the following we will simply refer to it as the regional ventilation, as impedance changes are assumed to be linearly related to the relative changes of aeration of the lung tissue.
If the computational model correctly resolves the regional phenomena within the lung parenchyma as a function of the pressure-controlled ventilation, the simulated EIT $Z^s (t)$ should provide similar information regarding the regional ventilation state of the lung as the bedside EIT $Z^r (t)$ acting as our reference.
At this point, it is important to note that identical results regarding regional ventilation would generally not be expected even without any model error. Discrepancies are inevitable, arising primarily from the unavoidable noise and epistemic uncertainty underlying the recorded clinical data, e.g., movement of attached electrodes, improper electrode contact, cardiac activity, or deviations of the lung from the assumed reference state. Additionally, the inverse problem underlying the EIT reconstruction is inherently ill-conditioned \cite{dorflerWhatInverseIllPosed2023}, imposing intrinsic bounds on the accuracy of EIT.

\subsubsection*{Preprocessing}

The measured voltage data $v^r$ obtained from bedside EIT has been recorded at a temporal resolution of 50 Hz, and was passed through a low-pass filter with a cutoff frequency of \res{1.4} Hz to remove spurious noise.
As a further preprocessing step, EIT reconstructions resulting from GREIT were normalized to the unit range $[0,1]$ and subsequently clipped to the lung masks, as the objective is to assess the accuracy of the predicted regional ventilation \emph{within} the lungs.
This clipping operation to the lung masks is clearly evident in Figure \ref{fig:eit_full_comparison_end_inspiratory}, only showing \emph{pulmonary pixels} within the lung masks.

\begin{figure}[h]
	\centering
	\ifmonolithicfigures
		\includegraphics{JAPPL_pv_data.pdf}
	\else
		\includegraphics[scale=0.61]{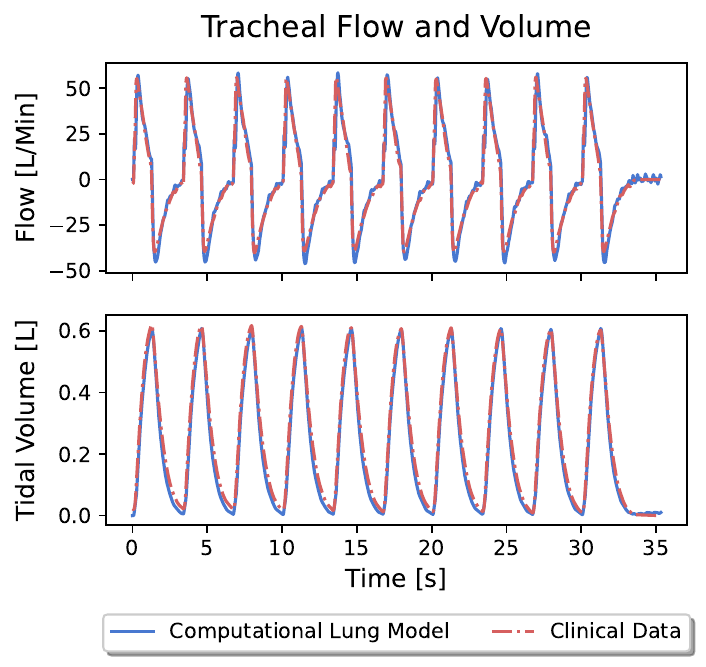}
	\fi
	\caption{Illustration of one exemplarily time-window underlying the validation, comprising 10 respiratory cycles. We compare the recorded flow rates and resulting tidal volumes exported from the mechanical ventilator (measured, clinical data) against the predictions obtained from the computational model. The waveforms are nearly identical.}
	\label{fig:pv_data_comparison}
\end{figure}

\subsubsection*{Data and Maneuver selection}

To mitigate the confounding impact of measurement noise on results, all ensuing comparisons are based on a continuous time window $t \in \left[ t_1, t_2 \right]$ comprising $10$ identical pressure-controlled ventilation cycles.
Time windows span approximately 30 seconds, and additionally contain a prolonged end-expiratory hold to establish a well-defined reference state required for difference-based EIT.

Exemplary ventilatory data associated with such a selected time is shown in Figure \ref{fig:pv_data_comparison}, showing both the measured data as well as corresponding model predictions.
While the agreement is of course expected given the calibration of the model on the waveform data and the constant ventilation settings, it nonetheless demonstrates that the model has successfully been attuned to the global, macroscopic characteristics of the respiratory system, defining a prerequisite for the validation of regional and localized predictions of the model.
Since the tidal volume remained fixed across all $10$ ventilation cycles, the lung repeatedly attains an identical ventilation state corresponding to a specific fraction of maximum observed end-inspiratory lung volume.
By averaging over multiple time instances corresponding to the same inflation state, we obtain a more robust identification of the regional ventilation with reduced impact of measurement noise and cardiac activity.

\begin{figure*}[t]
	\centering
	\ifmonolithicfigures
		\includegraphics{JAPPL_projected_ventilation.pdf}
	\else
		\includegraphics[scale=0.70]{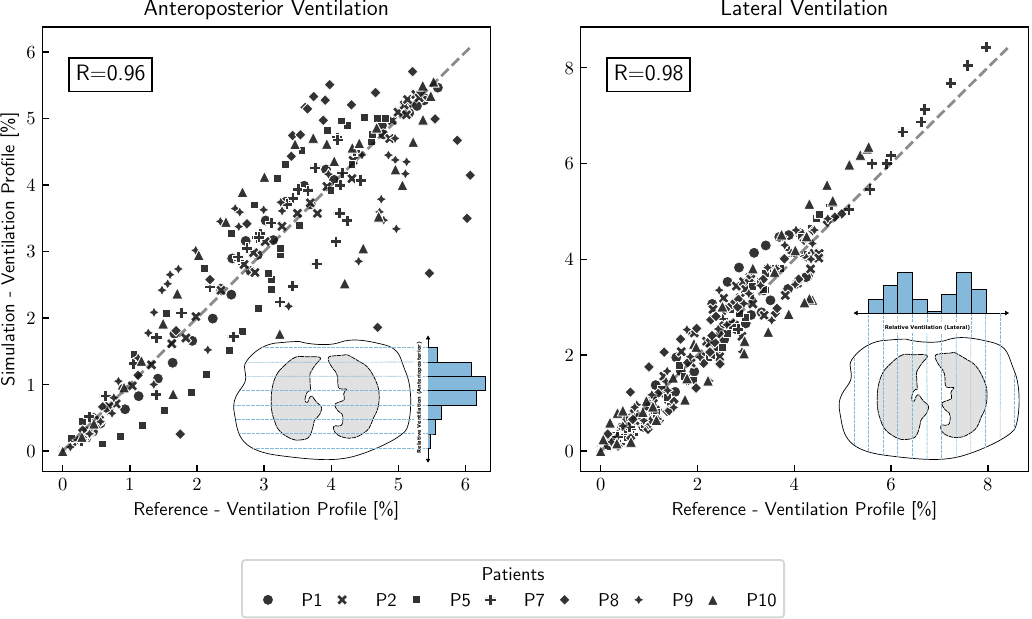}
	\fi
	\vspace{0.15cm}
	\caption{The two-dimensional end-inspiratory regional ventilation is projected (i) anterior to posterior and (ii) laterally left to right, yielding one-dimensional ventilation profiles. Data points indicate the fraction of ventilation in percent for anteroposterior or lateral slices of the patient, as indicated by the pictograms in the bottom right corners. The dashed line indicates the ideal unbiased one-to-one relationship of relative ventilation, with P1 to P7 indicating the attribution of data points to the N=7 patients.  }
	\label{fig:valdation_projected_views}
\end{figure*}

\subsubsection*{Quantitative Metrics}
We divide our discussion of validation results into three subsections, successively expanding upon the spatial and temporal resolution of the comparison between model-predicted and clinically measured regional ventilation
in each step.
First, we assess model agreement at the end-inspiratory state by projecting the two-dimensional ventilation distribution onto the anteroposterior and the lateral axis, respectively.
We chose this particular comparison because it has been previously employed in the validation of EIT against 4D-CT data \cite{thurkEffectsIndividualizedElectrical2017} and as such the known agreement between EIT and 4D-CT data essentially provides us with an upper limit or benchmark for the achievable agreement between EIT and model predictions.
In a second step, we directly compare ventilation distribution obtained from model predictions and measured voltage data at the end-inspiratory state at full available resolution of the EIT reconstructions, i.e., a pixel-wise comparison.
We note that for practical reasons, the end-inspiratory state has been defined as having reached $95\%$ of the tidal volume, i.e., $I(t) = 0.95$, with $I$ defined by the normalized cumulative sum of pixel intensities.
Lastly, we investigate the agreement between model predictions and EIT measurements across the entire breathing cycle to demonstrate that our model can also capture the dynamic behavior of the lung across the full tidal range.
To quantify the agreement of the regional ventilation derived from model predictions and point-of-care pulmonary lung monitoring in the ICU, we will primarily report the linear Pearson correlation $R$ and the root mean square error (RMSE) for all comparisons.
Please note that while the computational model of course yields predictions for the entire lung, the following discussions and results are always constrained to a two-dimensional transverse plane defined by the positioning of the EIT electrodes, as the available reference data (measured voltage data) does not permit reconstructions beyond this constrained plane.

\subsection{Regional Comparison at End-Inspiratory State }
\label{sec:validation_roi_and_derived}

\begin{figure*}
	\centering
	\ifmonolithicfigures
		\includegraphics{JAPPL_full_comparison_end_inspiratory.pdf}
	\else
		\includegraphics[scale=0.78]{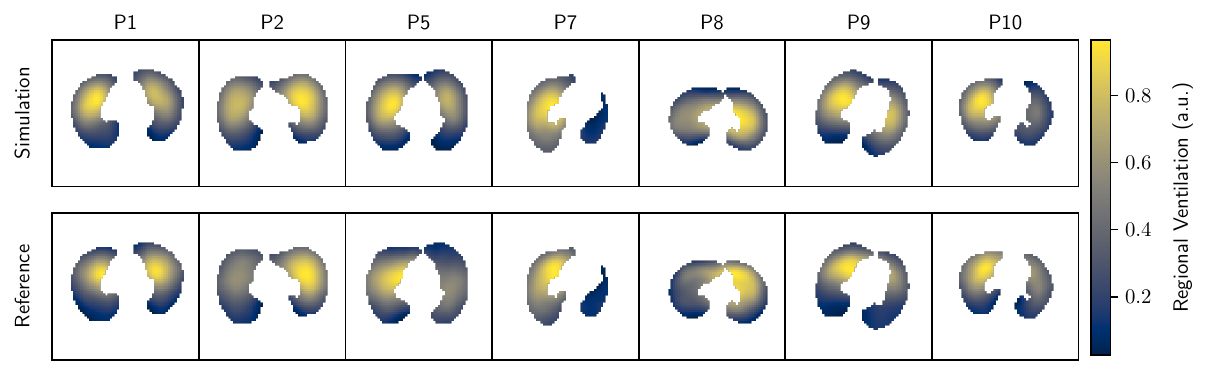}
	\fi
	\vspace{0.05cm}
	\caption{Regional ventilation at the end-inspiratory state for each of the seven patients. Predictions of the computational model (top row) are compared against clinical reference data (bottom row), indicating also the correlation coefficient $R$. The EIT images are clipped to within the patient-specific lung masks known from the CT (i.e., pulmonary pixels). }
	\label{fig:eit_full_comparison_end_inspiratory}
\end{figure*}

While we will strive to validate the full regional ventilation across the transverse plane, most prior published literature makes use of more coarse-grained and therefore more forgiving descriptions of the ventilation distribution such as the center of ventilation or a quadrant-based representation \cite{katayamaRegionalVentilationDynamics2024}.
In order to arrive at a metric that can be related to prior published work, we therefore initially - for the sake of comparability - also adopt a coarse-grained representation of the regional ventilation.
Recalling that $Z \in \mathbb{R}^{64 \times 64}$ defines the reconstructed EIT image at end-inspiration, we project the regional ventilation distribution $Z$ onto one-dimensional axes along the anteroposterior \emph{(ap)} and lateral \emph{(lat)} directions.
This implies a one-dimensional simplification of the ventilation distribution, where exemplarily, the anteroposterior (ap) ventilation profile is defined as
\begin{align}
	\text{{vp}}^{\text{ap}}_i = \frac{100 \cdot \sum_{j=1}^{64} Z_{ij}}{\sum_{i=1}^{64} \sum_{j=1}^{64} Z_{ij}}, \quad \text{for } i = 1, \dots, 64
\end{align}
The definition of the \emph{lateral} profile $v_j^{\text{lat}}$ follows analogously via summation over $i$. The result is a normalized one-dimensional attribution of fractional ventilation values along the specified anatomical axis, expressed as a fraction of total tidal ventilation within the transverse EIT plane. This approach employed in various studies \cite{hinzRegionalVentilationElectrical2003, frerichsReproducibilityRegionalLung2007, thurkEffectsIndividualizedElectrical2017} still retains relatively high spatial granularity along each axis while simplifying multidimensional comparison.
Figure \ref{fig:valdation_projected_views} illustrates the outcome of this analysis, aggregating data across $N=7$ validation subjects, comparing model-predicted and reference ventilation profiles plotted against the two axes.
With the dashed line representing a perfect one-to-one correspondence, the closer a data point to this line, the better the agreement of computational model and clinical data.
The symmetry of the data points with regards to this line moreover suggests a largely unbiased prediction of the computational model when compared against the clinical data. The outlier behavior of patient P8 most likely arises from the various complications associated with EIT as an imaging modality, and might for instance represent a mismatch of in-vivo electrode placement and numerically modeled electrode placement.
Since we employ a patient-specific torso geometry and mesh, it should also be noted that the individual projected ventilation profiles are intrinsically normalized with regards to the anatomical variability of individual patients (making the pooled representation of $N=7$ patients more comparable).
Importantly, when this same projection-based method was applied to validate EIT itself against 4D-CT in \cite{thurkEffectsIndividualizedElectrical2017}, the highest reported correlation for anteroposterior ventilation was $R=0.89$ (and similarly $R \approx 0.90$ when considering ventilation of quadrants \cite{katayamaRegionalVentilationDynamics2024}).
Our results thereby indicate that the agreement of the numerical model approaches the intrinsic resolution limits of EIT as determined by its own agreement with 4D-CT, which in turn implies evidence for the computationally predicted regional ventilation to the extent permitted by EIT and its intrinsic limitations.

\subsection{Comparison of End-Inspiratory Ventilation Distribution}
\label{sec:validation_direct_comparison}
In this section, we assess the agreement of the regional ventilation distribution for the \emph{full} spatial resolution of EIT (across a transverse plane), thereby expanding beyond the resolution limitation of the regions of interest in the previous section.
To this end, we obtain ventilation distributions using GREIT from simulated and measured voltage data at the end-inspiratory state at a resolution of $64 \times 64$ pixels.
The obtained reconstructions for the entire cohort are shown in Figure \ref{fig:eit_full_comparison_end_inspiratory}, with the top row illustrating the predicted ventilation distribution defined by the computational model and the bottom row showcasing the ventilation distribution obtained from the measured clinical voltage data.
Summarizing the results, Table \ref{tbl:correlation_inflated} presents the correlation $R$ and RMSE of pulmonary pixel intensity values in arbitrary units (AU). On average, we observe a correlation of \res{$R=0.82}$ and a (normalized) \res{$\text{RMSE}=0.14$}.

	\begin{table}[h!]
		\centering
		\setlength{\tabcolsep}{3pt}
		\begin{tabular}{l|ccccccc}
			\toprule
			Patient & 1      & 2      & 5      & 7      & 8      & 9      & 10     \\
			\midrule
			\midrule
			R       & $0.91$ & $0.94$ & $0.82$ & $0.92$ & $0.58$ & $0.82$ & $0.77$ \\
			\midrule
			RMSE    & $0.10$ & $0.10$ & $0.14$ & $0.12$ & $0.20$ & $0.16$ & $0.16$ \\
			\bottomrule
		\end{tabular}
		\vspace{0.20cm}
		\caption{Metrics reported for the validation cohort at end-inspiratory state.}
		\label{tbl:correlation_inflated}
	\end{table}

	Note that the RMSE, due to the mentioned normalization of EIT reconstructions to the unit range,
	attains an interpretable nature as the percentage of deviation, with the discrepancy between model-predicted and clinical reference ranging from \res{$10$} to \res{$20\%$}.
	Together with visual inspection of the regional ventilation distributions depicted in Figure \ref{fig:eit_full_comparison_end_inspiratory}, which for a subset of the patients approach visual indistinguishability, results demonstrate very good qualitative and quantitative agreement between model predictions and clinical reference data.

	\subsection{Comparison of Ventilation Distribution Across Full Tidal Range}
	\label{sec:validation_direct_comparison_full_tidal_range}

\begin{figure}[h]
		\centering
		\ifmonolithicfigures
			\includegraphics{JAPPL_ventilation_amplitude.pdf}
		\else
			\includegraphics[scale=0.58]{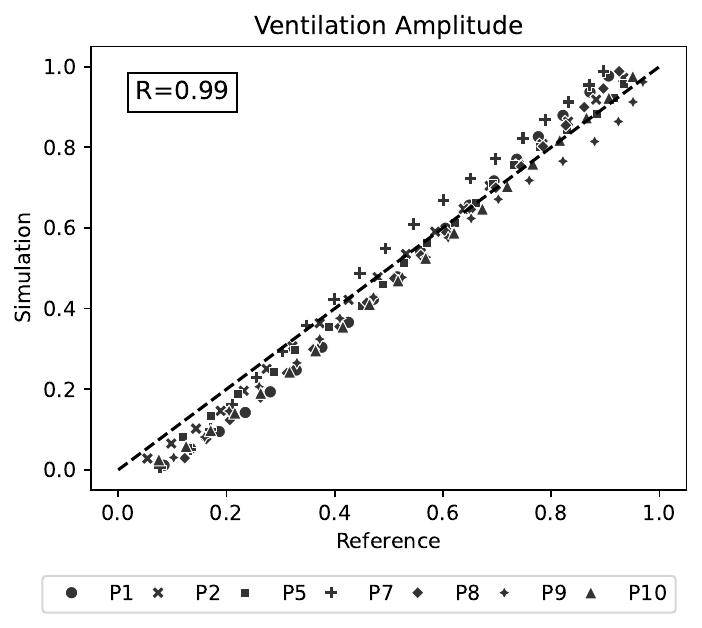}
		\fi
		\caption{Comparison of ventilation amplitudes derived from simulation and reference EIT across multiple inflation states. Each point represents the cumulative, unitless regional signal amplitude, normalized to the subject's tidal volume.}
		\label{fig:eit_validation_amplitude_correlation}
	\end{figure}

	The preceding discussion focused on assessing the accuracy of the model-predicted regional ventilation at the end-inspiratory state ($I=0.95$), at which the lung parenchyma experiences maximum inflation.
	To additionally study the dynamic model behavior, here we expand the previous discussion to the full tidal range occurring during ventilation.
	Since in such a setting we wish to disambiguate between regional distribution and net total change of aeration, we decompose both the model-predicted as well as reference ventilation distributions into the total amount of aeration $A(t)$, as well as its normalized regional ventilation $\hat{Z} (t)$, i.e., $Z (t) = A(t) \cdot \hat{Z} \left( t \right)$.
	By enforcing that $\hat{Z}$ has a unit-norm, it represents the regional ventilation distribution, while $A(t)$ becomes a function of overall pulmonary pixel intensities, corresponding to the total amount of aeration present.
	On the basis of this representation we can then decompose the task of validation into \emph{(i)} quantifying agreement of model-predicted and reference amplitudes $A$ (total amount of aeration), as well as \emph{(ii)} repeating our previous analysis of the regional distribution from section (\ref{sec:validation_direct_comparison}) for $\hat{Z}$ arising across multiple inflation states.
	For the ventilation amplitude $A(t)$ as a scalar quantity it is straightforward to proceed, with Figure \ref{fig:eit_validation_amplitude_correlation} illustrating the model's ability to provide an accurate and unbiased prediction of the change of cumulative aeration (evaluated for a discrete set of inflation states). Since $Z(t)$ is defined in terms of arbitrary units normalized here to the tidal volume, we plot dimensionless and normalized data points without any physical units.
	The minor bias discernible in Figure \ref{fig:eit_validation_amplitude_correlation} likely results from the increase of the signal-to-noise ratio towards the end-expiratory state, which non-symmetrically affects the noisy clinical voltage data compared to the electrodynamic simulation.
	While we already expected the model to correctly resolve the total amount of \emph{global} aeration, the agreement evident in Figure \ref{fig:eit_validation_amplitude_correlation} also demonstrates that the various steps detailed in Figure \ref{fig:concept_illustration_validation} - including electrodynamic forward simulation and inversion - all behave as expected, without introducing any notable distortion or significant bias.

	Having established correct predictions for the net total amount of aeration $A(t)$, we now turn our attention to assessing the accuracy of the model-predicted relative \emph{regional} ventilation for varying inflation states (spanning the entire tidal range).
	To this end, we repeat our previous analysis regarding correlation of pulmonary pixel intensities from the preceding section.
	The obvious change is that this time the comparison is applied to the normalized representations $\hat{Z}$, and repeated across multiple inflation states corresponding to $I = \left\lbrace 0.2, 0.4, 0.6, 0.8, 0.95 \right\rbrace$.

\begin{table}[h!]
	\centering
	\setlength{\tabcolsep}{3pt}
	\begin{tabular}{l|ccccccc}
		\toprule
		$I=$                 & $0.20$ & $0.40$ & $0.60$ & $0.80$ & $0.95$ \\
		\midrule
		R ($\varnothing$)    & $0.81$ & $0.81$ & $0.82$ & $0.83$ & $0.82$ \\
		\midrule
		RMSE ($\varnothing$) & $0.15$ & $0.15$ & $0.15$ & $0.14$ & $0.15$ \\
		\bottomrule
	\end{tabular}
	\vspace{0.10cm}
	\caption{Average metrics $R$ and RMSE for the regional ventilation across different ventilation states.}
	\label{tbl:metrics_dynamic}
\end{table}

With the resulting metrics averaged over the population ensemble listed in Table \ref{tbl:metrics_dynamic}, we observe that the model agreement remains largely unaffected and stable across all possible ventilation states comprising the full tidal range of mechanical ventilation.

\section{Discussion}
\label{sec:discussion}
In this study, we have provided systematic validation of a reduced-dimensional computational lung models, comparing numerical predictions of the regional ventilation against measured clinical reference data for mechanically ventilated ARDS patients.
By not just simulating the biomechanical processes within the lung during mechanical ventilation, but additionally also the precise mechanism by which the ventilation distribution has been inferred \emph{in vivo}, we ensured consistency and comparability between model predictions and clinical measurements (electrodynamic coupling).

Very good to excellent agreement between numerical predictions and reference clinical data was observed, even despite the pathophysiological inhomogeneities of the lungs and the known intrinsic limitations of the clinical data.
In addition to demonstrating agreement of anteroposterior and lateral projections of the ventilation state close to the resolution limitations of EIT, our computational model also yielded predictions largely congruent with the reference data even when considering the \emph{full} spatial resolution across the \emph{entire} dynamic range of mechanical ventilation.
Considering the complex nature of the clinical data as well as the required modeling assumptions underlying the electrodynamic simulation, it furthermore seems plausible that remaining observable discrepancies may at least partially be attributed to data limitations.
Exemplary, the evident anteroposterior shift or rotation for patient 8 and 10 in Figure \ref{fig:eit_full_comparison_end_inspiratory} might be explained by small inaccuracies in the modeling of the electrode placements in the electrodynamic simulation.
Even for cases exhibiting larger discrepancies between model predictions and clinical measurements, however, it seems plausible that the clinical information conveyed by model predictions concerning the lung is qualitatively very close to the measured EIT.

The computational lung model, therefore, provided correct predictions of the heterogeneous ventilation distribution and successfully accounted for the strong intra- and inter-patient variability of the pathophysiological lung state underlying ARDS.

Limitations of the validation results presented are defined by the available sample size ($N=7$) and otherwise largely coincide with the inherent limitations of EIT itself: \emph{(i)} restriction of the validation to a two-dimensional transverse plane, \emph{(ii)} potential errors or bias introduced in the reconstruction, \emph{(iii)} measurement noise and epistemic uncertainty associated with the clinical data.
In addition, both the validation and model construction are anchored to a single time point within the context of a pathophysiologically evolving disease. Regarding data requirements for patient-specific model construction, it is important to note that the acquisition of a chest CT scan typically involves both radiation exposure and clinical overhead. However, in cases of moderate to severe respiratory failure, such imaging is often already performed as a matter of clinical routine. Therefore, the model is intentionally designed not to introduce additional clinical burden or compromise patient safety.

    Beyond the availability of clinical data, our modeling choices also imply inherent limitations regarding the physiological processes that can be explicitly represented. In its current formulation, the model does not incorporate immunological, biochemical, or microscale biomechanical processes - such as surfactant dynamics, alveolar stability, or gas-liquid interface shifts - which are known to influence pulmonary mechanics in ARDS. More generally, the parenchymal tissue defines a multi-scale system of cascading complexity, with processes unfolding at spatial and temporal scales that generally exceed both the resolution limits of available clinical data as well as the envelope of feasible computational resources for large-scale personalized deployment. Our modeling strategy thus focuses on reproducing observed behavior at the tissue-scale by using representations that capture the aggregate effects of microscale processes, implicitly encoded through effective constitutive laws prescribed for alveolar clusters and informed from available clinical data (see Appendix \ref{appendix:A}).
    This stands in contrast to bottom-up multiscale models that seek to derive macroscopic behavior from fundamental physiological processes on the microscale, thereby shifting the model's resolution toward increased spatial and physiological complexity. Such models may incorporate, e.g., surfactant biophysics, surface tension dynamics, gas-liquid interface shifts, alveolar interdependence via shared septa, collagen-elastin fiber behavior or alveolar collapse thresholds. In principle, they can even be extended to include interactions with immunological or inflammatory mechanisms. Notable examples  would for instance be the incorporation of surfactant dynamics \cite{maSurfactantMediatedAirwayAcinar2020,maFulllungSimulationsMechanically2023} or - as continued development of the model presented here - the explicit representation of recruitment and derecruitment processes \cite{geitnerApproachStudyRecruitment2023, geitnerPressureTimedependentAlveolar2024}.
    While such extensions of the model offer greater explanatory depth and support for hypothesis testing or causal inference, they typically require more extensive parameterization, entail higher computational costs, and suffer from reduced robustness in patient-specific or data-limited settings. As a result, efforts in this direction remain challenging to calibrate and difficult to scale for clinical application.
    Our modeling approach, by contrast, deliberately aims to reproduce macroscopic behavior at the tissue-scale. It reflects a conscious trade-off between physiological fidelity and practical feasibility, enabling scalable analysis of regional lung mechanics while deferring fine-grained mechanistic interpretation to future studies or higher-fidelity nested models. Importantly, this trade-off is not determined by data availability or computational limits alone, as the foundational purpose of any model is to reduce reality to a manageable representation for a specific predictive task. In the clinical context of lung-protective ventilation, this task can be viewed through a decision-theoretic lens: from this perspective, the resolution of physiological processes is only relevant to the extent that it improves the basis for clinically actionable choices conditional on the available data and available computational resources (see full discussion in Appendix \ref{appendix:A}).
    This implies that physiological nuance meaningful in foundational research can introduce unnecessary or even counterproductive complexity in a clinical setting, if it does not improve decision-making in practice. In patient-specific applications, increased model detail must always be weighed against the feasibility of parameter identification, data availability, and real-time computational constraints. We therefore adopt the guiding principle that model complexity should serve clinical purpose, and consequently our modeling choices are governed by considerations of predictive utility, interpretability, and alignment with clinical use.
    Naturally, the model complexity spectrum is continuous. The eventual incorporation of more detailed mechanisms such as surfactant dynamics or alveolar collapse thresholds remains a highly worthwhile and promising direction for clinical application as well, in particularly as richer physiological datasets and imaging modalities become available. Towards this end, the  underlying model formulation is modular by design and can be extended to include additional physiological processes by introducing supplementary state variables and governing equations (as, e.g., the previously recruitment and decrecruitment processes \cite{geitnerApproachStudyRecruitment2023, geitnerPressureTimedependentAlveolar2024}). Thereby the foundational model structure defines a pathway for progressively increased physiological fidelity in the future, while being compliant with current constraints derived from clinical applications.

To our knowledge, and in alignment with regards to our previous statements on clinical feasibility, our results represent the first in-depth and quantitative validation study of patient-specific lung models providing detailed, regional information about the ventilation and mechanical state of heterogeneously damaged parenchymal lung tissue during mechanical ventilation.

In addition, our simulation approach is embedded into a comprehensive computational framework that includes automated image processing, model generation, and calibration.
In contrast to previous methodological proof-of-concept research \cite{rothComputationalModellingRespiratory2017, rothCouplingEITComputational2017}, which involved a lot of manual overhead for model creation and simulation setup, we can create simulation-ready models from images in minutes and deliver final results and analysis within an hour. As a result, our approach is fast enough to be applied in a clinical setting.

In closing, we seek to summarize and position the computational model within the broader landscape of pulmonary monitoring.
    As discussed in earlier sections, the dependence on thoracic CT imaging introduces certain temporal limitations compared to ultrasound- or EIT-based pulmonary monitoring. EIT, for example, offers continuous bedside measurements but is commonly restricted to a single transverse slice and provides only relative impedance-derived metrics. In contrast, the computational model - though initialized from a single time point - yields a volumetric, high-resolution reconstruction of regional lung mechanics across the entire parenchyma. Furthermore, once initialized, it requires no additional sensors or hardware and can operate non-invasively using data already available in standard clinical workflows for patients with moderate to severe respiratory failure.

Most notably, while we contrasted the model against EIT for the purposes of validation, it does not merely extend existing monitoring modalities but rather represents a different paradigm altogether. By introducing a physics-based, model-driven representation of the lung, it enables interpretability, predictive simulation, and mechanistic inference.
    Crucially, our model computes \emph{absolute}, time-resolved mechanical quantities—such as regional stress, strain, and mechanical power—that are interpretable, physiologically grounded, and directly comparable across patients. These quantities serve as key substrates for investigating the mechanistic drivers of VILI. Beyond quantifying total energy dissipation, the model enables decomposition of mechanical power into constituent components - such as elastic, resistive, and flow-associated contributions - allowing their individual roles in tissue damage to be examined. Recent work has emphasized that not all sources of mechanical energy are equally injurious and has proposed refined formulations to distinguish the damaging potential of different ventilatory mechanisms \cite{gattinoniVentilatorrelatedCausesLung2016, silvaPowerMechanicalPower2019,mariniWhichComponentMechanical2020}. Complementing this mechanistic breakdown, the spatial and temporal resolution of the model allows for identification of localized stress concentrations and energy hotspots, which are features increasingly suspected to underlie the initiation and regional propagation of VILI \cite{gaverMechanicalVentilationEnergy2025,hamlingtonAlveolarLeakDevelops2018,gottmanScalefreeModelAcute2024}.
    The spatial and temporal resolution offered by computational models provide a pathway for probing these causal relationships systematically, including the effects of ventilator settings and  tissue heterogeneity.
    Together, these capabilities open a path toward more mechanistically grounded, spatially resolved understanding of ventilatory harm.

In addition, by enabling the simulation of hypothetical or counterfactual ventilatory strategies, the model supports \emph{in silico} experimentation and intervention planning. Its generative capacity also opens avenues for data-centric applications: producing synthetic patient cohorts for training machine learning models \cite{phellanRealtimeBiomechanicsUsing2021}, augmenting limited datasets, or simulating outcomes in scenarios that are clinically infeasible to observe. Taken together, these features position the model not only as a diagnostic or monitoring tool, but as a scalable platform for advancing mechanistic understanding and data-driven decision support in critical care.

\section{Conclusion}
\label{sec:conclusion}
\begin{figure}[h]
    \centering
       \setlength{\belowcaptionskip}{0pt}
    \ifmonolithicfigures
        \includegraphics{JAPPL_clinical_application.pdf}
    \else
        \includegraphics[scale=0.41]{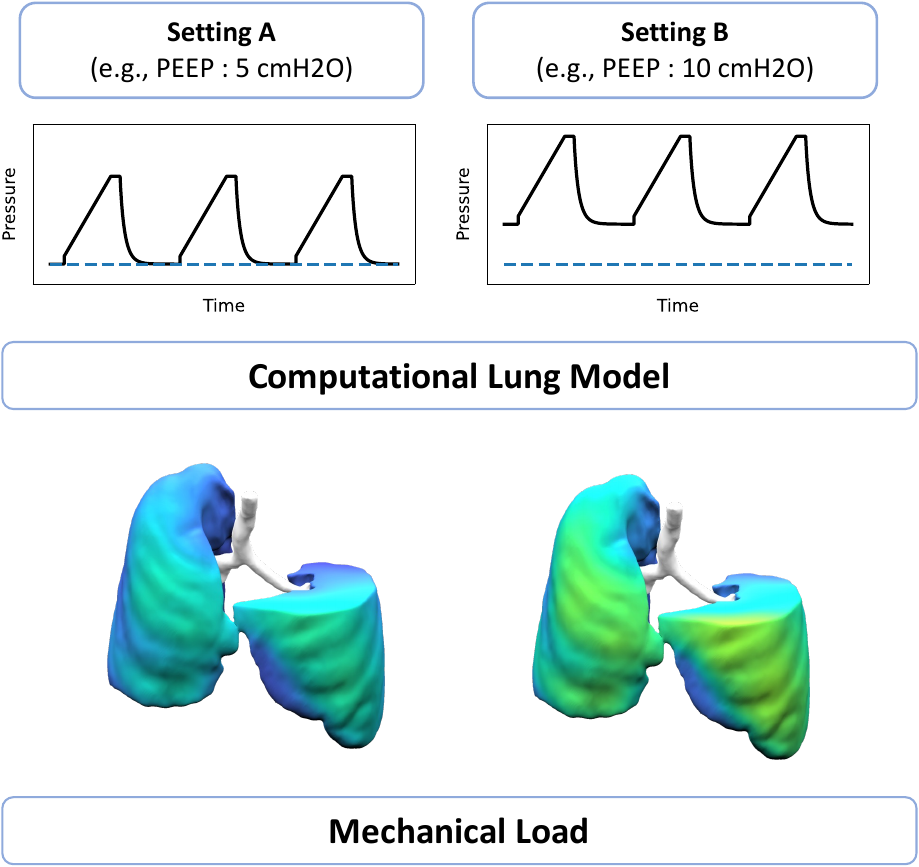}
        \vspace{0.20cm}
    \fi
    \caption{The computational lung model enables predictions regarding multiple hypothetical choices of the ventilation settings, yielding exemplarily the spatially resolved cumulative mechanical load over time as a function of the chosen PEEP.}
    \label{fig:clinical_application_setting_comparison}
\end{figure}

Having established the ability of the computational lung model to correctly resolve the regional biomechanical behavior of the lung parenchyma during mechanical ventilation, this opens up the perspective to link model predictions with assumed pathways underlying VILI.
By making use of highly resolved biomechanical models as employed in our work, one may introduce novel, model-based mechanical biomarkers to objectively quantify the overall regional ventilatory load for individual patients.
As indicated in Figure \ref{fig:clinical_application_setting_comparison}, this implies that one could use the computational model to \emph{(i)} predict how different hypothetical ventilation settings affect the lung and its ventilation, \emph{(ii)} compare the risk of injurious processes to the lung tissue and \emph{(iii)} ultimately choose an optimal patient-specific lung protective ventilation strategy, which is deemed optimal according to predicted biomarkers.

Towards this goal, future work will involve the validation of model predictions against 4D-CT data, which in contrast to EIT offers an improved spatial resolution at the expense of decreased temporal resolution.
Additionally, there is ongoing research to extend the computational lung model and its validation towards more complex physical and physiological phenomena, such as exemplarily resolving recruitment and de-recruitment phenomena \cite{geitnerApproachStudyRecruitment2023,geitnerPressureTimedependentAlveolar2024}.

\vspace{-0.05cm}

\clearpage
\backmatter

\subsubsection*{Funding}
This research received no specific funding.

\subsubsection*{Competing interests}
J.B., K.R.W., K.W.M., and W.A.W. hold shares of Ebenbuild.
All other authors declare no competing interests.

\subsubsection*{Ethics approval and consent to participate}
For this study, we analyzed data that was collected within the scope of the SMART-Study (DRKS-ID: DRKS00017151).
The study was approved by the Ethics Committee of the Medical Faculty of the Christian-Albrechts-Universität Kiel (Ethics Committee Number D 485/17) and carried out in accordance with the Helsinki Declaration.
Written informed consent was obtained from all patients or their legal representatives.

\subsubsection*{Consent for publication}
Not applicable.

\subsubsection*{Availability of data and materials}
All the data described and generated in the paper and SI is available upon reasonable request from the corresponding author.
Our tooling for the electrodynamic simulation and the EIT reconstruction, making use of EIDORS, has been open-sourced and is available online at \url{https://gitlab.com/xenotaph3/holoshed}.
The biomechanical simulations described in this paper were created using our own in-house codebase.
Unfortunately, the source code underlying the model generation and biomechanical simulation constitutes a proprietary asset which we are unable to share at the current time.

\subsubsection*{Authors' contributions}
T.B., J.B., I.F., D.S., K.W.M., and W.A.W. conceptualized and designed the research. T.B. planned and conducted the clinical study to collect the data. T.B., I.F., M.L., M.Lu., and A.S. collected and analyzed the clinical data. M.R. performed the simulation setup and runs.
M.R., M.Lu., and K.R.W. contributed to software development and model setup.
M.R., K.R.W., M.C.W, M.Lu., and J.B. did post-processing, data analysis, and visualization of simulation results.
M.R. and J.B. wrote the manuscript. All authors reviewed the
manuscript.

\subsubsection*{Acknowledgements}
T.B., I.F., and W.A.W. gratefully acknowledge the support from the Deutsche Forschungsgemeinschaft (DFG, German Research Foundation) in the project BE6526/1-1|WA1521/26-1.
  W.A.W. and M.Lu. gratefully acknowledge the support from BREATHE, a Horizon 2020-ERC-2020-ADG project (101021526-BREATHE).

\subsubsection*{Abbreviations}

\begin{center}
  \begin{tabular}{ r l}
    \textbf{ARDS}  & Acute Respiratory Distress Syndrome \\
    \textbf{AU}    & Arbitrary Units                     \\
    \textbf{CT}    & Computed Tomography                 \\
    \textbf{EELV}  & End-Expiratory Lung Volume          \\
    \textbf{EIT}   & Electrical Impedance Tomography     \\
    \textbf{GREIT} & Graz Consensus Reconstruction       \\
                   & Algorithm for EIT                   \\
    \textbf{PEEP}  & Positive End-Expiratory Pressure    \\
    \textbf{RMSE}  & Root Mean Square Error              \\
    \textbf{VILI}  & Ventilator-Induced Lung Injury      \\
  \end{tabular}
\end{center}

\clearpage
\renewcommand{\appendixname}{}
\begin{appendices}
  \section{Appendix}
\label{appendix:A}

\subsection*{Computational Lung Model}
\label{app:comp_lung_model}

The following provides a short introduction to the computational lung model, which has been developed over the last two decades and published in a number of peer-reviewed papers (as cited in the main text). It is intended as a service for the reader and strives to convey the basic concepts, without requiring detailed prior knowledge in the domain of computational physics and numerical modeling.

Before we delve deeper into how the model is constructed, let us briefly consider the physiological features and processes that the computational lung model has to resolve during the course of mechanical ventilation: for positive pressure ventilation, a driving pressure at the trachea leads to a non-zero transpulmonary pressure affecting the distension and inflation of the lung parenchyma with corresponding airflow and tidal volume.
Starting from the endotracheal tube during the inspiratory phase, air traverses in initially transitional flow down the larger airway structures and disperses into the lung parenchyma by following sequences of ever smaller self-similar asymmetric branching structures.
After 16 to 24 generations of conductive airways, the now fully laminar airflow reaches the respiratory zone, with the air subsequently dispersing into respiratory bronchioli, then alveolar sacs, finally reaching roughly 480 million individual alveoli where diffusion-based gas exchange occurs over a large surface area.
For the pathological lung parenchyma, this delicate and complex structure loses its relative homogeneity, with mechanical properties as well as resulting inflation and deflation cycles becoming increasingly heterogeneous in nature \cite{gattinoniRegionalPhysiologyARDS2017}.
Thereby, the already existing intrinsic intra-patient variability between healthy individuals (e.g., lung volume) increases further due to local pathophysiological variability introduced by ARDS, such as - exemplarily - inflammatory processes \cite{gonzalez-lopezLungStrainBiological2012} leading to consolidations or compression atelectasis in dependent regions.
The challenging situation faced by clinicians then lies in the complex manner in which the pathophysiological state of the lung parenchyma interacts with the forces exerted by the mechanical ventilator.
The resulting strain, stress, and overall trauma strongly depend on the \emph{specific} forcing pressure and flow applied at the trachea, which is determined by the clinicians choice of ventilator settings \cite{crookeIntracyclePowerDistribution2022, mariniWhichComponentMechanical2020}.
In order to predict and numerically capture the biomechanical behavior of the lung, it is therefore necessary to identify a model that can resolve the fundamental underlying physical phenomena subject to the particular physiological and pathophysiological state.

\subsubsection*{Patient-specific Model Construction}
\label{subsec:apx_model_construction}
To accomplish the goal of a \emph{patient-specific} model, the first step of model construction leverages machine learning and computer vision approaches to automatically segment and identify relevant physiological features from a thorax CT\footnote{Additional manual inspection and correction were performed for the validation cases, since the focus lies on assessing the model predictions}.
This step yields information on the exact shape, volume, and boundaries of the lungs, their subdivision into lobes along fissure lines, as well as the airway tree extending from the trachea, as shown in Figure \ref{fig:pipeline_model_construction}.
When reaching the resolution limit of the CT scan, the segmented airways are further extended by a space-filling airway tree growth algorithm \cite{tawhaiGenerationAnatomicallyBased2000} that continues the dyadic and self-similar branching structure within each lobe according to known statistical and morphological properties \cite{weibelMorphometryHumanLung1963}, yielding exemplarily the airway tree depicted in Figure \ref{fig:pipeline_model_construction}.
The use of these hybrid patient-specific/morphometric airway trees is based on the premise that beyond a certain structural size, it is no longer the specific geometry of individual airways that matter but rather the morphological properties of the airway tree.
The generational depth at which one truncates the growth of the artificial airway tree can be seen as a measure of model complexity - we consider up to 16 generations, reaching the threshold of the respiratory airways.
Beyond the terminal airways, the model does no longer consider individual structures in terms of airways or alveoli, and instead introduces models introducing an aggregate description of the compound behavior of bronchioli, alveolar sacs and alveoli - this can be regarded as lumped descriptions of the \emph{local} behavior of the respiratory zone.
For both airways as well as the respiratory zone the formulation we employ introduces reduced-dimensional models that characterize salient behavior at significantly reduced numerical cost. %
In consequence the entire process of model construction and numerical simulation results remains feasible to deliver personalized results within the hour, and all validation results where computed on a single-socket machine with a Intel(R) Xeon(R) Gold 6430 64-core processor and 512GB of RAM.

In the following, we provide a brief \emph{conceptual} introduction, discussing the reduced-dimensional elements for airways and lung parenchyma, which serve as model primitives and - once globally coupled - capture the behavior of the entire respiratory system.

\subsubsection*{Conducting Airways}

\begin{figure}[h]
    \centering
    \ifmonolithicfigures
        \includegraphics{JAPPL_0D_model_airway.pdf}
    \else
        \includegraphics[scale=0.85]{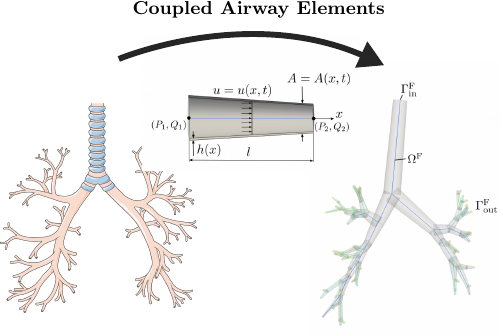}
    \fi
    \caption{The airway tree is reduced to coupled 0D airway elements, derived from the Navier-Stokes equation under simplifying assumptions. Illustration adapted with permission from Roth \cite{rothMultidimensionalCoupledComputational2016}. Copyright © 2016 Christian J. Roth. Adapted with permission. Additional element from Servier Medical Art (\url{https://smart.servier.com/}), licensed under CC BY 4.0.}
    \label{fig:airways_0d_model}
\end{figure}

In the simplest case, an individual airway segmented from the CT scan can be considered as an elongated tubular structure of a certain length $l_{aw}$ and radius $r_{aw}$ through which laminar, incompressible, and rotationally symmetric flow $Q$ occurs purely as a function of the forcing pressure differential $\Delta p$ as well as the resistance to the flow offered by the airway. This can be expressed as

\begin{align}
    Q = \frac{\pi \cdot r_{aw}^4}{8 \cdot \eta \cdot l_{aw}} \cdot \Delta p ~~,
    \label{Eq:poiseuille_flow}
\end{align}
with $\eta$ the dynamic viscosity of the fluid (Poiseuille flow).
In essence, this simplifies a complex three-dimensional fluid-structure interaction dependent on patient-specific geometrical features $(l_{aw}, r_{aw})$ into a highly simplified description. This is what - in abstract terms - we refer to an \emph{airway element} of the model.
Since we also know for each branching point of the airway tree that mass and momentum must be conserved, descriptions of individual airway elements can be coupled and assembled into a global system, implying that one can jointly determine time-dependent pressure and flow over the entire airway tree (see Figure \ref{fig:airways_0d_model}).
Adding increased complexity in the physical model of an airway element (compressibility of air, turbulence, non-rigid airways) will lead to increased complexity of the mathematical model defined by Eq. \eqref{Eq:poiseuille_flow}, but the conceptual idea remains unchanged. The bifurcations, which can be considered as junctions of airways, contribute additional equations that enforce conservation of mass and momentum and thereby couple the airway elements to each other.  \\

    In terms of model limitations, it is important to acknowledge that lumped 0D models are inherently unable to capture spatially dependent phenomena such as wave reflections and standing wave patterns. To this end consider for instance the Womersley number
    \begin{align}
        \text{Wo} = r_{aw} \sqrt{\frac{\omega \rho}{\mu}} ~,
    \end{align}
    which quantifies the ratio of unsteady inertial to viscous forces in oscillatory flow, where $r_{aw}$ is the characteristic radius, $\omega$ the angular frequency, and - as before - $\rho$ and $\mu$ correspond to the fluids density and its dynamic viscosity, respectively.
    Analysis of the resulting flow regime shows that for typical ventilation rates of adults (10-20 breaths per minute), viscous effects dominate, supporting the formation of stable, laminar velocity profiles. In contrast, high-frequency oscillatory ventilation (HFOV) as provided in neonatal care operates in a regime where unsteady effects become highly significant, thereby necessitating careful adaptation of the model to resolve pertinent physical phenomena  \cite{rothGasExchangeMechanisms2018, goderbauerComputationalModellingAssisted2022}.
    Similarly, the Reynolds number
    \begin{align}
        Re = \frac{\rho u D}{\mu},
    \end{align}
    describes the ratio of inertial to viscous forces and thereby informs the onset of turbulence. When the Reynolds number suggests non-laminar behavior, additional losses due to turbulence can be incorporated through nonlinear resistive elements.
    For the problem under consideration, however, neither turbulence, compliance of airways, nor unsteady inertial effects (e.g., pressure-flow phase lag) were significant enough to justify additional modeling complexity.
    As our principal objective lies in the validation of the computational model against  clinical data, here - as elsewhere - we have foregone the introduction of additional model complexity, if the limitations and relatively coarse resolution of the EIT reference data did not permit to draw inference on its validity (as any consistent model selection approach in a data-driven setting will always consider explanatory power against model or parameter complexity.)

Once the flow of air has traversed through the maximum number of resolvable airway generations (up to 16), we model the local behavior of the lung parenchyma in the surrounding area of each terminal airway, as detailed in the following section.

\subsubsection*{Alveolar Clusters}

The lung parenchyma beyond the conducting airways - comprised of respiratory bronchioli and alveoli - is modeled by alveolar clusters that capture the \emph{locally} homogenized behavior of a small subregion of the lung parenchyma.
The name of this substitute model element - alveolar cluster - derives from it capturing the aggregate behavior of a local \emph{cluster} of several thousands or tens of thousands of alveoli.
Intuitively explained, alveolar clusters define local exit-compartment models attached to terminal airways coupled via flow and pressure.
It acts as a substitute model for a local ensemble of small-scale respiratory bronchioli and alveoli in the vicinity of the terminal airway, emulating the local behavior of the lung parenchyma, complying and/or counteracting the airflow through terminal airways through various static, dynamic, and viscous effects. In \cite{rothCouplingEITComputational2017} originally an Ogden-type material with additional viscous effects was proposed for the alveolar clusters, implying the following functional dependence of transpulmonary pressure $p_{tp}$ on acinar volume $V$ and flow $Q$:

\begin{align}
    p_{tp} = \frac{V_0}{V} \frac{\kappa}{\beta} \left( 1 - \left( \frac{V_0}{V} \right)^\beta \right) + \frac{\eta}{V_0} Q
    \label{Eq:odgen_type}
\end{align}

\begin{figure}[h!]
    \centering
    \ifmonolithicfigures
        \includegraphics{JAPPL_ac_model.pdf}
    \else
        \includegraphics[scale=0.80]{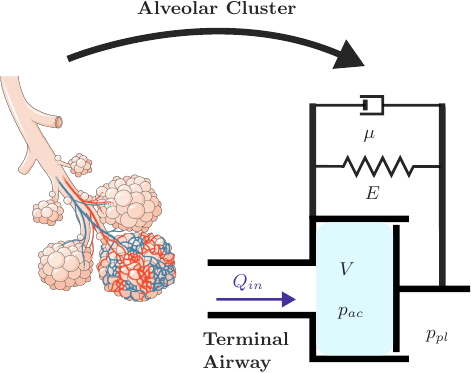}
        \vspace{0.35cm}
    \fi
    \caption{
        The respiratory zone of the lung parenchyma, comprised of bronchioli and alveoli, is modeled as alveolar clusters, acting as an exit compartment for the terminal airways. Mechanically equivalent behavior can be expressed in terms of simple elements such as, e.g., springs and dampeners. Illustration makes use of element from Servier Medical Art (\url{https://smart.servier.com/}), licensed under CC BY 4.0.
    }
    \label{fig:ac_0d_model}
\end{figure}

Here $\kappa$ and $\beta$ denote constitutive parameters of the lung parenchyma, while $\eta$ defines the viscosity.
Due to the limitations of EIT and the relatively small tidal volumes, in our case the alveolar clusters have been further simplified to Kelvin-Voigt elements.
A schematic illustration of a reduced-dimensional Kelvin-Voigt alveolar cluster element at the distal end of the conducting airway tree is given in Figure \ref{fig:ac_0d_model}, relating pertinent physical phenomena to mechanically equivalent behavior, i.e., springs or dampening elements, counteracting the inflow (or outflow) of air $Q$. Its defining constitutive parameters - in mechanical analogy - are spring stiffness $E$, reference volume $V_0$, and volume-normalized viscosity $\mu$. Spatial variability of constitutive parameters and reference volumes are informed by the variability of Hounsfield attenuation values recovered from the CT, as will be detailed in our section on model calibration.
The forcing pressure differential experienced by each alveolar cluster corresponds to the transpulmonary pressure and arises as the pressure differential between alveolar and pleural pressure $p_{pl}$.
More formally, the alveolar clusters capture the local behavior of the lung parenchyma through a viscoelastic model relating strain and stress within the lung tissue \cite{modestoialapontClinicalImplicationsRheological2019, guerinRegionalLungViscoelastic2021, birzleExperimentalCharacterizationModel2018, birzleConstituentspecificMaterialBehavior2019, birzleViscoelasticNonlinearCompressible2019, birzleCoupledApproachIdentification2019}.
Similarly to the discussion for airways, increasingly complex physical phenomena giving rise to increasingly complex rheological behavior can be modeled by expanding upon the prototypical spring and damper elements depicted in \ref{fig:ac_0d_model}, allowing to construct for instance generalized Maxwell elements.

    It is important to emphasize that the elements used to represent regional lung behavior reduce the description to an effective viscoelastic behavior \cite{faffeLungParenchymalMechanics2009}. They therefore reproduce the observed mechanical responses of lung tissue - such as pressure-volume hysteresis or viscoelastic relaxation - without explicitly modeling the underlying biophysical mechanisms at the microscale. As outlined in our discussion in detail already, this defines a deliberate choice, and contrasts with fully mechanistic or first-principles models, seeking to shift the truncation of complexity to incorporate increasingly sophisticated phenomena on ever smaller length-scales. In the following we shall briefly discuss in greater depth the rationale and thereby the associated question of suitable model complexity in the context of clinical applications.

\subsubsection*{Model Complexity}

    When considering specific modeling choices and a suitable cutoff point within the physiological complexity cascade, we should recall that the underlying purpose of the computational model is to aid in the identification of lung-protective ventilation settings. This goal can be formalized as a Bayesian decision problem: let $ s \in \mathcal{S}$ denote the true physiological state of the patient, $a \in \mathcal{A}$ a possible action (e.g., ventilator setting), and $u(s, a)$ a utility function encoding the clinical benefit or harm of choosing action $a$ in state $s$. Given observed clinical data $\mathcal{D}$, the intractable optimal decision is abstractly defined as

\begin{align}
    a^*(\mathcal{D}) = \arg\max_{a \in \mathcal{A}} \mathbb{E}_{s \sim \pi(s \mid \mathcal{D})}[u(s, a)] ~,
\end{align}

where $\pi(s | \mathcal{D})$ is the posterior distribution over states inferred from data. When repeatedly adjusting actions over time $a_t$ for a dynamically evolving $s_t$, this essentially defines a dynamic treatment regime.  In practice, the physiological state is however not directly observable and the utility $u$ also only defined indirectly in terms of complex patient-ventilator interactions. Once constructed and calibrated according to available patient data, the computational model defines an implicit and approximate point-estimate of the patient's internal state projected onto the finite model resolution subspace $\mathcal{S}_{\text{model}}$, summarizing the relevant mechanical and structural features subject to the assumptions of the model class. Subsequent clinical reasoning or simulation can then be understood as operating on this inferred state to predict the effect of alternative actions, i.e. ventilator settings, by virtue of $u(s,a)$. Within the context of the model this involves time-dynamical simulations where the action $a$ defines generally periodic boundary conditions applied at the trachea, dictated by e.g., specific choices of PEEP and driving pressure.

    The resulting implication is that actions only depend on aspects of the state $s$ which influence the utility $u(s, a)$ and are also inferable from available clinical data $\mathcal{D}$ via the posterior $\pi(s | \mathcal{D})$. This observation is closely related to the existence of an equivalence relation $s_i \sim s_j \iff u(s_i, a) = u(s_j, a)\ \forall a \in \mathcal{A}$, partitioning the state space into equivalence classes of decision-relevant physiological states. The set of such equivalence classes forms a reduced space $ \mathcal{S}_\text{eff}$, which contains exclusively the features of $\mathcal{S}$ relevant to decision-making. Consequently, only the information in $\mathcal{S}$ that survives this projection influences optimal action selection. Any model detail surpassing the equivalence class within $\mathcal{S}_\text{eff}$ or rendered unidentifiable by the limitations of the clinical data $\mathcal{D}$ can generally not be expected to improve clinical decision-making.

    In this sense, both the clinical task as well as the resolution of the available clinical data $\mathcal{D}$ act as an information filter, restricting the effective resolution at which model predictions remain decision-useful. Adding more detail - e.g., by resolving finer spatial scales or including additional physiological mechanisms - may increase overall physiological fidelity of the model, but does not necessarily reduce decision regret unless these additions improve prediction of the decision-relevant features. If computational complexity and feasibility of realtime inference on available hardware is considered as well, additional considerations apply, implying that one should define the model as a pareto-optimal choice w.r.t. physiological fidelity and computational feasibility.
    Overall this justifies a design philosophy that prioritizes model resolution aligned with decision utility and subject to feasibility constraints, rather than exhaustive physiological representation.

\subsubsection*{Numerical Discretization}

Numerical discretization of the coupled alveolar cluster and airway descriptions yields a differential algebraic system of equations (DAE), which we solve iteratively by making use of suitable time-integration methods \cite{kunkelDifferentialalgebraicEquationsAnalysis2006} \textemdash \,  i.e., we simulate the state of the lung parenchyma over time, as a function of time-varying pressure or flow applied by the mechanical ventilator (depending on the ventilator mode).
In contrast to a simplistic one-compartment model, which has exactly one flow rate, one volume, and one pressure, depending on the chosen generational depth of the airway tree, the model-associated state vector $x(t) \in \mathbb{R}^n$ comprises $n=10^4$ to $n=10^6$ degrees of freedom, corresponding to time-dependent and spatially resolved flow, volume, pressure, strain and stress distributed over the entire lung parenchyma.
In particular, the various state variables of each alveolar cluster as well as the flow through the airway tree define a vector of unknowns $x(t) \in \mathbb{R}^n$, for which the solution is implicitly defined by the residual $F : \mathbb{R}^{2n + 1} \to \mathbb{R}^n$ as a function of the state vector $x(t)$

\begin{align}
    F \left( x(t), \dot{x}(t), t ; \theta \right) = 0
    \label{eq:DAE}
\end{align}

which in conjunction with a suitable $x_0 = x(t=0)$ defines an initial value problem (IVP) driven by either pressure or flow boundary conditions applied to the trachea (corresponding to either pressure-controlled or volume-controlled ventilation).
Please note that the specific dependence of the DAE on the parameters $\theta$ and the corresponding inference problem will be discussed in subsequent separate subsection on model calibration.

\subsubsection*{Pleurae, chest wall, and diaphragm}

An aspect not discussed in detail so far is that the behavior of the lung parenchyma during mechanical ventilation of course also depends on the surrounding pleural pressure.
The pressure in the pleural space and its spatial variability are defined by several aspects \cite{silvaRegionalDistributionTranspulmonary2018}, such as the compliance of the thoracic cage counteracting volumetric changes of the lung, sternocostal articulation, pleural interface, and intra-abdominal pressure.
In addition, hydrostatic pressure increases towards dependent regions conditional on gravity and patient positioning (e.g., prone, supine), and of course, any spontaneous breathing efforts by the patient result in muscular activity modulating the pleural pressure.
While the patients enrolled in the study were sufficiently sedated to disregard inspiratory efforts as a meaningful contributor to pleural pressure variations,  gravitational forces as well as elastic recoil of the chest wall are accounted for by means of a suitable external pressure boundary condition acting on the alveolar clusters. Pertinent patient-specific parameters relating to the chest-wall are identified and isolated during the calibration process using esophageal pressure measurements, as will be discussed in further detail.
The pleural pressure boundary condition also encompasses a hydrostatic pressure component that depends on the known positioning of the patient and on the weight of the lungs as determined from the CT image, performing a density and volume analysis to account for the spatial variability of the pleural pressure. Using the lung and lobar geometry as well as Hounsfield attenuation values from the CT, a hydrostratic pressure field $p_{hydro} (x,y,z)$ can be determined as a function of the varying density $\rho (x,y,z)$ within the lung parenchyma, which defines the spatial perturbation of the pleural pressure as a function of the weight of the enclosed organ. Correspondingly, a change in patient positioning (e.g. prone/supine) leads to different forces exerted on the lung parenchyma, impacting in turn the strain and regional ventilation.

\begin{draftBlue}
    \subsubsection*{Model Calibration}

    The computational model is intrinsically dependent on the patient-specific anatomy, primarily via geometric effects (e.g., lobar segmentation and airway topology). Additionally, however, the differential algebraic equation system \eqref{eq:DAE} exhibits dependence on a set of parameters $\theta$ associated with both regional and global descriptors of biomechanical behavior.
    The fundamental objective is to infer parameters $\theta \in \Theta$ that both yield consistent macroventilatory behavior (i.e., reproduce the observed waveform data from the ventilator) and consistently resolve volumetric and constitutive effects of the patient-specific physiology and pathophysiology (regional lung mechanics). To this end, in the following we will outline in detail
    \begin{itemize}
        \item the unknown parameters $\theta$ of the model
        \item the clinical data available for calibration
        \item consistency requirements
        \item inference and optimality
    \end{itemize}
    While the airway elements in principle could also contribute to $\theta$, under current assumptions their impact on $x(t)$ is entirely governed by known geometrical features or a-priori known morphological characteristics. In consequence, the only reduced-dimensional elements with unknown parameters contributing to $\theta$ are given by the alveolar clusters as well as the equivalent lumped-effect model of the thoracic cage defining the pleural condition.

    \begin{description}
        \item[Alveolar Clusters:] Given the previously adopted Kelvin-Voigt constitutive behavior, we assume independently varying properties of reference volumes $V_{0}^{(n)}$ and elastance values $E^{(n)}$ (as indicated by the additional presence of $n$, denoting the $n$-th alveolar cluster). In contrast, the volume-normalized viscosity $\eta$  as appearing in Eq. \eqref{Eq:odgen_type} is currently defined globally and shared among ACs. For a total of $n=1, ..., N$ alveolar clusters this implies the parameter set
              \begin{align}
                  \theta_{ac} = \left\lbrace \left\lbrace  E^{(n)}, V_{0}^{(n)} \right\rbrace_{n=1}^N , \mu \right\rbrace
              \end{align}
              comprised of individual reference volumes, elastances, as well a singular global viscosity value.
        \item[Pleural Condition:] The transpulmonary pressure experienced by each AC depends on the spatially varying hydrostatic pressure field, which in turn is informed by the CT scan. The only parametric dependence however is in the \emph{global} constitutive pressure-volume relationship, for which a model identical to the ACs was adopted:
              \begin{align}
                  \theta_{pl} = \left\lbrace V_{0}^{\text{pl}}, E^{\text{pl}} \right\rbrace
              \end{align}
    \end{description}

    This implies that our pertinent set of parameters is defined by joint contributions $\theta = \left\lbrace \theta_{ac}, \theta_{pl} \right\rbrace$ of alveolar clusters and thoracic cage, with the precise dimensionality of $\theta \in \Theta$ itself being patient-specific as a function of overall alveolar clusters $N$. More generally this provides a good opportunity to remind the reader that for the entire discussion, both data and inference is naturally always conditional on a \emph{specific} patient, albeit not made explicit in notation to avoid undue notational obfuscation.

    Pertaining to the clinical data, we recall that the waveform data defines a temporal sequence $\mathcal{W}$ over the given time-interval $\mathbb{T}$ comprising both pressure measurements as well as flow measurements obtained by the mechanical ventilator:

    \begin{align}
        \mathcal{W} = \left\lbrace p_t^{\text{aw}}, p_t^{\text{eso}}, Q_t^{\text{aw}} ~ \middle| ~ t \in \mathbb{T} \right\rbrace
    \end{align}

    This data can be understood to predominantly hold information on macroventilatory features of the respiratory system, which in the conventional clinical settings would be described approximately by means of - for instance - static and dynamic compliance.
    The second crucial data point available for each patient is the thoracic CT at a known pressure $p_{\text{ct}}$, encoding \emph{regional} information on the \emph{radiodensity} of the parenchymal tissue in terms of Hounsfield values. If we introduce $\Omega \subset \mathbb{R}^3$ to denote the spatial lung domain, then the CT can be understood as a voxel-wise scalar field $\mathcal{H} : \Omega \rightarrow \mathbb{H}, \quad x \mapsto \mathcal{H}(x)$ mapping any $x \in \Omega$ to its corresponding Hounsfield attenuation value $\mathcal{H} (x)$. This implicitly also defines a patient-specific empirical distribution $p(h) \approx |\Omega|^{-1} \sum_{x \in \Omega} \delta(h - \mathcal{H}(x))$, which is shaped by the underlying ventilation and structural heterogeneity (consider, e.g., focal vs. non-focal ARDS), and thereby even without spatial attribution represents information on the pathophysiological state.
    To further our discussion we will also need to introduce the average Hounsfield value \emph{within} the lungs or subregions thereof as $\overline{h} = \mathbb{E}_{p(h)} \left[ h \right]$, as well as consider the prevalence of various gaseous, liquid and solid phases within the lung. To this end we follow a well-established approach \cite{gattinoniWhatHasComputed2001} and linearly interpolate between fully gaseous and liquid state to define a gaseous fraction $\gamma_{\text{gas}} \in [ 0, 1 ]$

    \begin{align}
        \gamma_{\text{gas}}(h) =  \text{sat}_{[0,1]} \left( \frac{h - H_{\text{water}}}{H_{\text{gas}} - H_{\text{water}}} \right)
        \label{eq:gas_frac}
    \end{align}

    , with $H_{\text{gas}} = -1000$ and $H_{\text{water}} = 0$. It follows from Eq. \eqref{eq:gas_frac} that we may ascertain the absolute amount of aeration present within any subdomain $\omega \in \Omega$ by virtue of the integral

    \begin{align}
        V_{\text{gas}}^{\omega} = V_{\text{gas}}(\omega) = \int_{\omega} \gamma_{\text{gas}} \left( \mathcal{H} ( x)  \right) dx ~ ,
    \end{align}

    as well as of course the density and weight of any subregion $\gamma$. With both the parameters and available clinical data well-defined, we can now turn our attention towards the actual calibration of the computational model.
    Instead of permitting arbitrary results for the parameters $\theta$ and simply attempting to match the observed waveform data $\mathcal{W}$, we a-priori posit several constraints: \\

    \begin{enumerate}[(i)]
        \item \textit{Global volumetric coherence:} The model-predicted static gaseous volume within the respiratory system (ACs and airways) must match the global observed gaseous volume $V_{\text{CT,gas}}$ encoded in $\mathcal{H}$ at the associated known airway pressure $p_{\text{ct}}$ applied during acquisition of the CT.
        \item \textit{Local volumetric coherence:} As each alveolar cluster defines a lumped-effect model of a certain subregion of the lung parenchyma $\Omega^{(n)}$, the relative gaseous volumes $V_{\text{gas}}^{(n)}$ as defined by $\gamma_{\text{gas}} (h)$ and the alveolar cluster domain $\Omega^{(n)}$ must be matched by the model.
        \item \textit{Physiological sufficiency:} Global macroventilatory respiratory features are defined by the cumulative contributions of ACs. To reduce complexity and provide the simplest possible example, if $c^{(n)}$ denotes the static compliance of the $n$-th alveolar cluster, then the global static lung compliance follows as $C_{\text{lung}} =\sum_{n=1}^N c^{(n)}$. This implicit summation constraint defines an affine hyperplane in compliance space, but permits infinitely many equally valid configurations - leading to both identifiability issues and permutation invariance across clusters. We therefore posit that properties of alveolar clusters may not be individually chosen, but that the local structure must uniquely be determined by the Hounsfield attenuation values of the acinar cluster domains $\Omega^{(n)}$ as well as the ventilatory waveform data $\mathcal{W}$, which defines the combined clinical data at our disposal.
    \end{enumerate}
    \leavevmode\newline

    The requirements \emph{(i)} and \emph{(ii)} imply a global coupling and a-priori constraint on the parameters. To comply with the principle articulated in \emph{(iii)}, we introduce a low-dimensional, context-aware constitutive map $\phi_{\xi} : \mathbb{H} \times \Xi \rightarrow \mathbb{E}$, where the Hounsfield attenuation values associated with an alveolar cluster serve as the context to inform local pathophysiological conditions.
    Conditional on the parameters $\xi \in \Xi$ this therefore defines a mapping from observed radiodensity $\mathcal{H} \in \mathbb{H}$ to the set of all elastance values $\left\lbrace E^{(n)} | n = 1..., N \right\rbrace \subset \mathbb{E}$, imposing both a physiological prior as well as a regularizing constraint.
    Concerning the structure of $\phi_{\xi}$ we assume that increased prevalence of parenchymal tissue and fluid will lead to increased resistance to volumetric deformations, implying that elastance should increase monotonically with attenuation values.
    Furthermore, we can impose desired limit-case behavior on $\phi_{\xi}$, such as e.g. the elastance diminishing to zero as attenuation values approach the gaseous domain $H_{\text{gas}}$. While this provides reasonable constraints and boundary conditions, the suitable choice $\phi_\xi$ is neither uniquely defined nor a-priori obvious, and should be subject to further future investigations conditional on the availability of highly resolved 4D-CT reference data. It is however clear from theoretical considerations that the constitutive map $\phi_\xi$ should exhibit scale-consistency or resolution-invariance, implying that the additional sub-partitioning of parenchymal tissue and introduction of further ACs in conjunction with $\phi_{\xi}$ yields consistent coarse-grained behavior.
    Subject to all these considerations and following previously published approaches \cite{rothCouplingEITComputational2017} we introduce as baseline assumption a first-order approximation, which encodes linear perturbations of the elastance $E^{(n)} = \xi_0 + \xi_1 ( h_{\text{ac}}^{(n)} - \overline{h} )$ arising from deviations from the mean Hounsfield attenuation, with $h_{\text{ac}}^{(n)} =  | \Omega^{(n)} |^{-1} \int_{\Omega^{(n)}} \mathcal{H}(x) dx$.

    Since the patients under consideration are subject to pressure-controlled ventilation, it follows that we use the airway pressure as a boundary condition and require matching predictions of the remaining waveform data $\tilde{\mathcal{W}}$ comprised only of tracheal flow and esophageal pressure. With the condensed set of parameters $\hat{\theta}$ resulting from the constitutive map $\phi_{\xi}$ and a-priori articulated constraints \emph{(i) - (iii)}, we introduce the loss over observable waveform data

    \begin{align}
        \hat{\theta}^* & = \arg \min \mathcal{L} \left( \hat{\theta} \right) \nonumber                                                                \\
                       & = \arg \min \left| \left| \hat{\mathcal{W}} - \hat{\mathcal{W}}_{\text{DAE}} \left( \hat{\theta} \right) \right| \right|_2^2
        \label{Eq:CalibrationObjtFct}
    \end{align}

    which we seek to minimize w.r.t. $\hat{\theta}$ subject to the previously outlined constraints. As the patients in question are ventilated via pressure-controlled ventilation, the loss function is defined, more specifically, by the misfit of model predictions against measured tracheal flow and esophageal pressure. In the context of discretely sampled data this yields

    \begin{align}
        \mathcal{L}(\hat{\theta}) = \sum_{t \in \mathbb{T}} \bigg[
         & \phantom{+} \left( \hat{Q}^{\text{aw}}_t(\hat{\theta}) - Q^{\text{aw}}_t \right)^2\nonumber \\
         & + \left( \hat{p}^{\text{eso}}_t(\hat{\theta}) - p^{\text{eso}}_t \right)^2
        \bigg]
        \label{eq:CalibrationObjtFct}
    \end{align}

    where it should be noted that the esopheageal pressure $p_t^{\text{eso}}$ is not directly defined for the computational model, as generally speaking equality of esophageal pressure and pleural pressure does not hold. Instead we assume that they are identical up to a constant, scalar offset parameter $\delta_{\text{pl-es}}$ which is subsumed within $\hat{\theta}$, and essentially acts as a latent variable.
    The actual numerical implementation makes use of a separate two-stage approach that first identifies global macroventilatory features, and subsequently uses the full-scale model to obtain $\hat{\theta}^*$ which minimizes the misfit defined by Eq. \eqref{eq:CalibrationObjtFct} and is in compliance with our postulated consistency requirements.
\end{draftBlue}

\section{Appendix}
\label{appendix:B}

\subsection*{Electrical Impedance Tomography}
\label{appendix:eit}

In order to employ Electrical Impedance Tomography for pulmonary monitoring during mechanical ventilation \cite{bachmannElectricalImpedanceTomography2018, tomicicLungMonitoringElectrical2019a}, an electrode belt is positioned around the thorax of each patient, with electrical stimulations  - i.e., driving current - being injected at high temporal resolution according to predefined and quickly alternating stimulation patterns.
Since the spatially variable impedance of the lung parenchyma depends on its aeration, the resulting potential field and the corresponding measured complex-valued voltage signals at the electrode belt are necessarily not only a function of the intrathoracic structure, but also depend on the regional ventilation distribution within the lung.
By quickly cycling through stimulation patterns and measuring the potential field at the remaining receiving electrodes, time-dependent information can be aggregated on the regional ventilation of the lung. Using a 32-electrode belt attached to an Elisa $800$ ventilator with integrated EIT, at each time-step complex-valued voltage data $v \in \mathbb{C}^{1024}$ is obtained at a sampling rate of 50 Hz (the dimensionality is defined by the number of electrodes as well as stimulation and measurement patterns). Both injection as well as measurement pattern follows a $(0,5)$ pattern, i.e. injection and measurement occurs with $4$ in-between electrodes, with an injected current of $5 \cdot 10^{-3}$ Ampere.
While there are certain practical complexities, measuring this clinical voltage data corresponds to well-established clinical processes already widely in use on ICUs during mechanical ventilation.
The subsequent reconstruction of the regional ventilation from the voltage data defines an \emph{inverse} problem, and - similarly to the discussion of the electrodynamic simulation \eqref{sec:electrodynamic_simulation} - is predicated upon the existence of a forward model, which essentially relates measured voltages with injected stimulation patterns and the conductivity of the lung parenchyma.

\subsubsection*{Forward Problem}

The behavior of electrodynamic fields propagating across the torso is known to be characterized by Maxwell's equations, which for our purposes - under several simplifying assumptions - can be reduced to a single, scalar elliptic partial differential equation on the domain of the patient's torso, i.e. for all $x \in \Omega_{torso}$

\begin{align}
    \text{div} \left( \sigma (x) \cdot \nabla \Phi (x) \right) = 0
    \label{eq:eit}
\end{align}

we relate the tissue conductivity $\sigma (x)$ and the resulting potential field $\Phi (x)$.
For the sake of our discussion it is not relevant to examine the specific structure of this equation, nor the specifics of the boundary conditions which have been omitted from explicit representation.
Instead we simply observe that both the conductivity $\sigma (x)$ and the potential field $\Phi (x)$ resulting from the currents injected at the electrodes spatially vary over the torso, and that their interdependence is defined by fundamental physical laws mathematically expressed by Eq. \eqref{eq:eit}. In the context of the simulated voltages, as previously detailed, the corresponding conductivity $\sigma (x)$ was informed by highly resolved reference simulations carried out for the microscopic structure of the lung parenchyma \cite{rothCorrelationAlveolarVentilation2015}.
Using well-established numerical discretization techniques such as Finite Elements \cite{wooFiniteelementMethodElectrical1994} this equation can be solved to obtain resulting simulated voltage, i.e., one computes the potential field across the torso conditional on the conductivity field and the stimulating currents applied at the electrodes (\emph{forward problem}).
In order to solve the forward problem we again rely on EIDORS \cite{adlerEIDORSCommunitybasedExtensible2005}, which is also employed for the solution of the inverse problem using GREIT (\emph{Graz consensus Reconstruction algorithm for EIT}).

\subsubsection*{Inverse Problem}

The fundamental goal of Electrical Impedance Tomography is however to traverse the \emph{inverse} route, i.e., to identify the conductivity field $\sigma (x)$ over the lung parenchyma which can adequately explain a set of observed voltage measurements.
This conceptually implies to minimize the misfit or maximize the likelihood of model predictions with regards to observed voltage data, generally subject to additional prior information or additional regularization (\emph{inverse problem}).
For pulmonary monitoring via EIT, we follow the established and generally more stable approach of \emph{difference} EIT, where regional ventilation is inferred on the basis of \emph{perturbations} of the voltage data from an end-expiratory reference state.
This behaves favorable, because unknown but constant obfuscators - such as the intrathoracic structure and variations of contact impedances of electrodes - do not enter directly into the reconstruction problem.
As previously mentioned, for the reconstruction  we employ GREIT \cite{adlerGREITUnifiedApproach2009}, as it is a well-established and widely used algorithm with an available open-source implementation \cite{adlerEIDORSCommunitybasedExtensible2005}.
The outcome of the reconstruction is generally a two-dimensional regional ventilation map varying over time, which may be represented as a two-dimensional $Z$, resolving the regional ventilation across a transverse cut-slice of the patient's thorax between the 4th and 5th intercostal space  (see Figure \ref{fig:ct_slices} and ventilation maps in Figure \ref{fig:concept_illustration_validation}).
As such, GREIT maps voltage data to two-dimensional intensity images. For the resolution we employ for our validation purposes, $Z$ is given by a $64 \times 64$ grid of pixels, which we may represent as $Z \in \mathbb{R}^{64 \times 64}$.  Mathematically, this is a square matrix comprised of real-valued numbers. GREIT therefore defines a mapping $\mathbb{C}^{1024} \mapsto \mathbb{R}^{64 \times 64}$, with voltage measurements $v$ giving rise to the EIT image $Z$.
In this EIT image, pixel intensity relates to an increase of inverse conductivity, and thus a larger degree of local aeration of the lung parenchyma.
For a more in-depth discussion of EIT for respiratory monitoring as well as the specifics of the inverse problem setting we refer the reader - due to the inherent scope and complexity - to the appropriate literature for further information \cite{frerichsChestElectricalImpedance2017,
    bachmannElectricalImpedanceTomography2018, adlerElectricalImpedanceTomography2021}.
For our purposes, we simply conclude that conditional on voltage data (be it simulated or measured) the inversion underlying EIT yields an image $Z$ carrying information on the local aeration and regional ventilation of the lung across a two-dimensional transverse plane.
By conducting the identical inverse reconstruction approach both for measured clinical voltage data as well as for simulated voltage data resulting from the computational lung model, we can compare predictions of regional ventilation against reference data.
While the computational lung model permits predictions beyond merely regional ventilation (such as e.g. local strain), the validation centers around the regional ventilation because it is the only primary information than can be derived directly from EIT without necessitating further assumptions and approximations.

\subsubsection*{Electrodynamic Coupling}

The biomechanical model yields predictions of the regional ventilation and does not intrinsically represent electrodynamic properties of the lung. In order to enable a comparison against bedside EIT, it is therefore necessary to map computational model predictions to alterations of bioimpedance. In order to define this mapping we make use of previously executed highly-resolved three-dimensional elastodynamic reference simulations, which determined the impedance of the lung parenchyma on a microscopic scale as a function of its relative aeration and tortuosity \cite{rothCorrelationAlveolarVentilation2015}.
Denoting the relative gaseous fraction within the tissue as the filling factor $\text{FF} = V_{\text{air}} / V_{\text{tissue}}$, it was found that the mean effective resistivity follows a linear relationship

\begin{align}
    \overline{\rho}_{\text{eff}} = \frac{\overline{\tau}}{\sigma_{\text{alv}}} ( \text{FF} + 1)
    \label{eq:filling_factors}
\end{align}

with approximately constant mean tortuosity $\overline{\tau} = 1.71$ and alveolar conductivity $\sigma_{alv} = 0.7228 ~ \Omega^{-1} \text{m}^{-1}$.
Since the computational lung model yields predictions of the filling factors, we can precisely mirror the stimulation patterns and electrode placements of bedside EIT and subsequently - on the basis of Eq. \eqref{eq:eit} - carry out additional electrodynamic simulations, to obtain \emph{simulated} voltage data in complete equivalence to the measured clinical voltage data. Please note that $\overline{\rho}_{\text{eff}}$ is independent of the choice of a specific strain measure and merely depends on the relative prevalence of gaseous content within the lung parenchyma.

\subsubsection*{Technical Details}

We close this section with some further technical details concerning EIT, which warrant explicit mention.
First, patient-specific torso and lung geometries used for the generation of simulated voltages are also used for the solution of the inverse problem underlying EIT, enabling reconstructions with a higher degree of fidelity compared to population-averaged torso geometries generally in use for bedside EIT \cite{thurkEffectsIndividualizedElectrical2017}. However, any tissue outside the lungs was simply assumed as homogeneous, i.e., neglecting varying conductivity of, e.g., bones and muscles.
Secondly, in the context of the validation effort it has to be observed that the employed reconstruction algorithm follows a heuristic approach and has several non-uniquely defined parameters that inevitably arise due to the necessity to regularize the ill-conditioned inverse problem arising from Eq. \eqref{eq:eit}, thereby implicitly affecting the results.
There does not exist an established generic manner of setting these parameters, as they are generally highly problem-dependent and will vary as a function of the specific characteristics of the measured clinical data (e.g, number and placement of electrodes, electrical and movement noise level, improper and detached electrodes).
In absence of definitive 4D-CT data, we have followed the most commonly employed approach of determining reconstruction parameters on the basis of qualitative assessment of reconstructions, with identical reconstruction parameters used across the entire patient cohort.
Likewise, reconstruction parameters of GREIT are chosen identical for both simulated as well as measured voltages with exception of a small discrepancy in the noise figure, due to the lack of noise affecting the simulated data.

\end{appendices}

\bibliography{bibliography}

\end{document}